\documentclass[preprint,12pt]{elsarticle}

\usepackage{amssymb}
\usepackage{lipsum}

\usepackage{lineno}
\usepackage{etoolbox,refcount}
\usepackage{multicol}
\usepackage{amsmath,amsthm,amsfonts,amssymb,amscd}
\usepackage{hyperref}
\usepackage{natbib}
\usepackage{cleveref}
\usepackage{rotating}
\usepackage{tabularx}
\usepackage{protosem}
\usepackage{color}
\usepackage{xcolor}
\usepackage[scr]{rsfso}

\newcommand{\RoMag}{\textsc{RoMag}}
\newcommand{\red}{\textcolor{red}}

\journal{International Journal of Heat and Mass Transfer}

\begin{document}

\begin{frontmatter}



\title{Thermovelocimetric Characterization of Liquid Metal Convection in a Rotating Slender Cylinder}


\author[first,UCLA]{Yufan Xu}
\author[UCLA]{Jewel Abbate}
\author[UCLA]{Cy David}
\author[HZDR]{Tobias Vogt}
\author[UCLA]{Jonathan Aurnou}
\affiliation[first]{organization={Princeton Plasma Physics Laboratory},
            addressline={100 Stellarator Rd}, 
            city={Princeton},
            postcode={08540}, 
            state={NJ},
            country={USA}}

\affiliation[UCLA]{organization={Department of Earth, Planetary, and Space Sciences, University of California, Los Angeles},
            addressline={595 Charles E Young Dr E}, 
            city={Los Angeles},
            postcode={90095}, 
            state={CA},
            country={USA}}
            
\affiliation[HZDR]{organization={Institute of Fluid Dynamics, Helmholtz-Zentrum Dresden-Rossendorf},
            city={Dresden},
            postcode={01328}, 
            country={Germany}}
            
\begin{abstract}
Rotating turbulent convection occurs ubiquitously in natural convective systems encompassing planetary cores, oceans and atmospheres, as well as in many industrial applications. While the global heat and mass transfer of water-like rotating Rayleigh-B\'enard convection is well-documented, the characteristics of rotating convection in liquid metals remain less well understood. In this study, we characterize rotating Rayleigh-B\'enard convection in liquid gallium (Prandtl number $Pr \approx 0.027$) within a slender cylinder (diameter-to-height aspect ratio $\Gamma = D/H = 1/2$) using novel thermovelocimetric diagnostic techniques that integrate simultaneous multi-point thermometry and ultrasonic Doppler velocity measurements. This approach experimentally reveals the formation of a stable azimuthal wavenumber $m = 2$ global-scale vortical structure at low supercriticality. We propose that enhanced wall modes facilitated by the slender cylinder geometry interact with the bulk flow to create these large-scale axialized vortices. Our findings extend results from the previous $Pr \sim 1$ studies across various cylindrical aspect ratios. In particular, we find evidence of a different scaling for wall mode precession frequency that possibly exists in liquid metal, offering new insights into the coupling effects in low-$Pr$ rotating convective turbulence.

\end{abstract}




\end{frontmatter}




\section{Introduction}
\label{intro}



Rotation and buoyancy-driven convection coexist in many astrophysical systems and industrial engineering applications. Rotating Rayleigh-B\'enard convection (RC) is a canonical, mathematically well defined model for studying buoyancy driven flows that are characterized by heating from below, cooling from above, and subject to rotation \cite{chandrasekhar1961hydrodynamic}, which allows direct comparisons between numerical simulation and experimental results \cite{madonia2021velocimetry,stevens2013heat}. Rapidly rotating flows are fundamental to understanding rotationally dominated quasi-geostrophic dynamics in the Earth's outer core \cite{gillet2011rationale,calkins2015multiscale,schaeffer2017turbulent,aubert2017spherical,calkins2018quasi,horn2022elbert,abbate2023rotating}, oceans, atmospheres \cite{mcwilliams2006fundamentals}, and other planetary fluid systems such as the underground oceans of icy moons \cite{schoenfeld2023particle}. Rotating convection physics is also applicable in the processing of materials, such as casting \cite{wei2008centrifugal} and solidification for crystal growth \cite{weber1990stabilizing,nikrityuk2006numerical,chen2010fluid,muller2012crystal}. It helps in achieving more uniform temperature distributions and can affect the microstructure of materials. Moreover, in chemical reactors and mixers, rotating convection can enhance mixing and heat transfer, improving reaction rates and uniformity \cite{visscher2013rotating}. 


Rotation inhibits the vertical transport associated with convection and the Coriolis force stabilizes convective instabilities, thus delaying the onset of convection in the bulk of the domain \cite{chandrasekhar1961hydrodynamic}. Four essential dimensionless control parameters for RC are the Prandtl number $Pr$, the Rayleigh number $Ra$, the Ekmen number $Ek$, and the aspect ratio $\Gamma$. The Prandtl number describes the thermo-mechanical properties of the fluid,
\begin{equation}
    Pr = \frac{\nu}{\kappa},
\end{equation}
where $\nu$ is the fluid's viscous diffusivity, $\kappa$ is the thermal diffusivity. In liquid gallium, $Pr \approx 0.027$ for most of our experiment cases which have a mean temperature at $\approx 40^\circ \mathrm C$, and it decreases slightly with increasing temperature. For the case with the highest heating power at $\sim 2$kW, $Pr \approx 0.024$. 
The Rayleigh number characterizes the buoyancy forcing relative to thermoviscous damping, 
\begin{equation}
   Ra = \frac{\alpha_T g \Delta T H^3}{\nu \kappa}.
\end{equation}
where $\alpha_T$ is the thermal expansion coefficient, $g$ denotes the magnitude of the gravitational acceleration, $\Delta T$ is the temperature difference between the bottom and the top of the fluid layer, and the characteristic length scale is given by the layer height $H$.
The Ekman number describes the ratio of viscous and Coriolis forces,
\begin{equation}
    Ek=\frac{\nu}{2 \Omega H^{2}},
\end{equation}
where $\Omega$ is the angular rotation rate of the system. The aspect ratio $\Gamma = D/H $ is a diameter-to-height ratio, and $\Gamma = 0.5$ is fixed for this study. Another important parameter is the convective Rossby number, which is the ratio between buoyancy-fueled inertia and Coriolis,
\begin{equation}
    Ro_C= \sqrt{\frac{\alpha_Tg\Delta T}{4 \Omega^2 H}} = \sqrt{\frac{RaEk^2}{Pr}}.
\end{equation}
The convective Rossby number is argued to control gross transitions in RC behavior in numerous studies \cite[e.g.,][]{gastine2014zonal,aurnou2017cross,aurnou2020connections,horn2015toroidal}.

Plane-layer RC has been well studied in asymptotic theories and direct numerical simulations (DNS) \cite[e.g.,][]{malkus1954heat,rossby1969study,julien1996rapidly,grossmann2000scaling,ahlers2009heat,favier2014inverse,guervilly2014large,rubio2014upscale,kunnen2016transition}. Asymptotic quasi-geostrophic numerical models adopt a novel approach by modifying the governing equations to address the conditions of very rapid rotation ($Ro_C \lll 1$) \cite{sprague2006numerical,julien2012statistical,julien2016nonlinear,plumley2016effects}. These asymptotically reduced models, though simplified, are vital as they aim to replicate behaviors pertinent to large-scale planetary flows that current geophysical models cannot directly simulate.

Laboratory experiments, on the other hand, are intrinsically limited by the geometry and aspect ratio of the container. Cylindrical geometry is often adopted in RC experiments, which is often motivated to simulate a high-latitude fluid parcel \cite{aurnou2015rotating,cheng2015laboratory,cheng2018heuristic,cheng2018heat,kunnen2021geostrophic}. Thermal forcing is imposed at both end-blocks at the top and bottom. While it is clear that convective dynamics differ between polar and equatorial regions \cite{glatzmaier1995three,simitev2005prandtl,gastine2023latitudinal}, this setup provides a controlled environment to further dissect these differences and their impact on global dynamo processes. 

Laboratory studies involving parameter scaling often use slender cylinders with an aspect ratio $\Gamma = D/H < 1$ to achieve high Rayleigh numbers $Ra$ and low Ekman numbers $Ek$, since both parameters are highly dependent on the vertical length scale ($Ra \propto H^3$, and $Ek \propto H^{-2}$). However, as the aspect ratio decreases, the physical presence of the sidewall exerts a growing dynamic influence on flow dynamics. Precessing wall modes, characterized by retrograded alternating vertical flows at the circumference of the cylinder, develop at the onset of convection \cite{herrmann1993asymptotic,zhang2009onset,horn2017prograde}. The wall modes improve overall heat transfer at low supercriticality and interact with the bulk dynamics, with increasing significance as the aspect ratio decreases, or equivalently, as the boundary curvature of the sidewall increases \cite{herrmann1993asymptotic, madonia2021velocimetry, de2023robust, zhang2024wall}. 



Consequently, dynamics generated due to the presence of a sidewall have non-negligible effects on the heat and momentum transfer of RC's bulk flow. Therefore, the small aspect ratio containers with sidewalls may pose a serious challenge to laboratory experiments designed to investigate bulk dynamics in RC. However, to date, RC studies have been focused on water and water-like experiments in which $Pr \gtrsim 1$. It remains unclear if the wall-bulk interaction persists and if large-scale vortical structures can happen in lower-viscosity fluids, such as liquid metals.  

Liquid metal is a low-$Pr$ electrically conducting fluid. As decreasing $Pr$ results in higher resolution requirements for the numerical simulations, turbulent RC with core-like material properties is expensive for the current numerical models to resolve. Most experiments and simulations of rotating convection typically use fluids with $Pr \gtrsim 1$, such as air ($Pr \sim 0.7$) or water ($1.75 \leq Pr \leq 13.5$) 
\cite[e.g.,][]{zhong1993rotating,julien1996rapidly,zhong2010heat,king2012heat,stevens2013heat,horn2014rotating,cheng2015laboratory,kunnen2021geostrophic}. In lieu of the difficulties in low-$Pr$ numerical simulations, laboratory experiments in liquid metal have realistic material properties where $Pr \sim 10^{-2}$. 


Because liquid metals are opaque, it is impossible to obtain optical high-resolution velocity field measurements by Particle Image Velocimetry (PIV) as demonstrated in water experiments \cite[e.g.,][]{kunnen2010experimental,aujogue2018experimental,madonia2021velocimetry}. It is thus challenging to identify lengthscales and structures using the cross-correlation method over the horizontal cross-section of the cylinder. Nevertheless, we have demonstrated in this study that large-scale vortical flow structures could be characterized by combining multiple 1D UDV velocity measurements and analytical models of potential vortices. 

The main goal of this study is, for the first time, to investigate thermal and velocity dynamics of the sidewall, bulk, and wall-bulk interactions in a turbulent liquid metal RC system with an aspect ratio $\Gamma = 1/2$. We conducted a total of four RC experiments on UCLA's $\RoMag$ device. Two sets of RC experiments were carried out near the onset of wall mode and oscillatory bulk convection at $\sim20$ revolutions per minute (RPM), which yield $Ek \sim 5 \times 10^{-7}$. Another two sets of RC experiments implemented more substantial thermal forcings at a slightly slower rotation $\sim 12$ RPM, and $Ek \sim 8 \times 10^{-7}$. 

\Cref{theory} summarises onset predictions of convection and phenomena in liquid metal RC. \Cref{method} describes our experimental setup and analytical tools for characterization flow behaviors. \Cref{wallmode} presents our thermovelocimetric results of the prominent wall modes. \Cref{quad} describes our characterization of a quadrupolar cyclone generated from wall-bulk interaction. \Cref{disc} and \Cref{conclusion} are discussion and conclusion for this study. 

\section{Liquid metal RC behavior regimes in a cylinder} 
\label{theory}

Here we summarize the onset predictions of various modes in liquid metal RC. Specifically, since infinite plain layer theory for rotating convection has been studied extensively, we focus on asymptotic onset predictions with lateral boundaries to represent a cylindrical geometry. A more general review of different modes in RC has been discussed previously \cite{horn2017prograde,aurnou2018rotating,kunnen2021geostrophic}. Geometrical confinement has a stabilizing effect on both oscillatory and stationary rotating convective instabilities, leading to higher critical Rayleigh numbers for convective onsets \cite{goldstein1993convection,goldstein1994convection}.


\subsection{Oscillatory mode}
Liquid metal rotating convection can produce a wide spectrum of phenomena \cite{aurnou2018rotating, vogt2021oscillatory}. Most significantly, the thermal convective instabilities in liquid metal rotating convection develop as overstability in the form of thermally driven oscillations \cite{zhang1997thermal} instead of the quasisteady columnar flow (stationary mode) in $Pr\gtrsim0.68$ fluids \cite{chandrasekhar1961hydrodynamic, niiler1965influence,homsy1971asymptotic}. These modes are oscillating helical structures that align with the rotation axis, transporting energy across the fluid bulk.  

The critical Rayleigh and horizontal length scale of oscillatory mode at the onset in liquid metal are functions of $Ek/Pr$. This is distinguished from steady rotating convection, which is only dependent on $Ek$. For $Pr\ll1$ fluid in the horizontally infinite plane,
\begin{equation}
    Ra_o^\infty \approx 17.4 \left(\frac{Ek}{Pr}\right)^{-4/3},\quad \ell_o^\infty \approx 2.4 \left( \frac{Ek}{Pr} \right)^{1/3},    
\end{equation}
where $Ra_o^\infty$ is the critical Rayleigh number and $\ell_o^\infty$ is the critical horizontal lengthscale. The oscillation frequency normalized by rotational frequency at the onset is estimated to be
\begin{equation}
    \widetilde f_o^\infty \equiv \frac{f_o^\infty}{f_\Omega} \approx 4.8 \left( \frac{Ek}{Pr} \right)^{1/3},    
\end{equation}
where $f_\Omega = 1/\tau_\Omega = \Omega/(2\pi)$. Here $\tau_\Omega$ is defined as the rotation period, or the time for the system to rotation once. However, far from infinite plane, our experiment was conducted in a finite geometry, a more accurate prediction is needed to interpret and compare with the experiment results. Zhang \& Liao \cite{zhang2009onset} provided asymptotic predictions in the limit of small $Ek$ and $Pr$ that account for the finite geometry, assuming the curvature of the cylinder can be negligible, the sidewall is thermally insulated, and top and bottom boundaries are isothermal and no-slip. Minimizing their equations (4.21) and (4.22) results in more precise estimates of the Rayleigh number $Ra$, oscillation frequency, and modal structure compared to predictions for an infinite layer \cite{horn2017prograde}.  

\subsection{Wall modes}
The presence of the side boundaries affects the onset of the rotating convection in liquid metal. Lateral boundaries can release rotational constraints and therefore destabilize the fluid. Wall mode is a convective instability that develops through a Hopf bifurcation. It is a wave-like structure with upwelling/downwelling patches and travels in the retrograde direction around the circumference of the tank \cite{herrmann1993asymptotic,zhang2009onset,zhang_liao_2017,grannan2022experimental}. 

Wall modes are predicted at low $Ek$ to first develop at \cite{zhang2009onset}
\begin{equation}
    Ra_W \approx 31.8 Ek^{-1}+46.6 Ek^{-2/3}.
\end{equation}
The azimuthal wavenumber is 
\begin{equation}
    m_W \approx \Gamma (3.03-17.5 Ek^{1/3}),
\end{equation}
and the onset frequency, normalized by rotation frequency, yields
\begin{equation}
    \widetilde{f_w} \approx 131.8 \frac{Ek}{Pr} - 1464.5 \frac{Ek^{4/3}}{Pr}.
\end{equation}
For simplicity, these predictions are written in approximation form. We calculated these predictions using proper forms \cite{zhang_liao_2017}, leading to less than $0.1\%$ error from the published results. Furthermore, following the asymptotic equations\cite{zhang2009onset}, we calculated the normalized frequency for oscillatory modes with our experiment setting, $\widetilde{f_o}$. These two frequency predictions are used as default in the following sections. Furthermore, we estimated the horizontal wavenumbers of these modes. At $Ek \approx 8\times 10^{-7}$, $Pr = 0.027$, $\Gamma = 0.5$, and assuming a simplest vertical wavenumber $n = 1$, the wall modes onsets at $Ra_w = 3.228\times 10^7$ with an azimuthal wavenumber $m = 1.43$; the bulk oscillatory mode onsets at $Ra_w = 4.248\times 10^7$ with an azimuthal wavenumber $m = 1$, and a radial wavenumber $k = 4$. Frequency prediction for each case is reported in the first two columns of \cref{tab:fft}.



\subsection{Wall-bulk interactions}
The presence of a boundary zonal flow (BZF) located near the sidewall of a $\Gamma \sim \mathcal O(1)$ cylinder is a key feature of $Pr \gtrsim 1$ rotating convection \cite{de2020turbulent,zhang2020boundary,zhang2021boundary,wedi2021rotating,de2023robust,ecke2023turbulent}. The boundary zonal flow emerges from wall modes \cite{favier2020robust, ecke2022connecting} and extends into the fluid interior as aspect ratio decreases \cite{de2023robust, zhang2024wall}. BZF exits from the onset of wall mode convection up to the transition to buoyancy-dominated flows at $Ro \approx 2$. 

Madonia et al. (2021) \cite{madonia2021velocimetry} investigated rapidly rotating convection with stereoscopic particle image velocimetry in water ($Pr = 5.2$) and in a $\Gamma = 1/5$ cylinder. Madonia et al. (2021) \cite{madonia2021velocimetry} found wall mode emanate jets into the bulk and set a $m = 2$ quadrupolar vortex in horizontal motion, using an orientation-compensated mean vorticity field at supercriticality $Ra/Ra_s^\infty = 47$. 

The quadrupolar vortex has also been found in numerical studies \cite{de2023robust,zhang2024wall}. De Wit et al. \cite{de2023robust} conducted direct numerical simulation with a small and a larger aspect ratio container, which yields $m = 1$ and $m = 2$ wall mode, respectively. As a result, in the $m=1$ wall mode case, the flow forms a quadrupolar vortex with two pairs of cyclones and anti-cyclones. In the wider cylinder with $m = 2$ wall mode cases, the flow forms eight smaller alternating vortices along the azimuth near the outer radius, and the jets have less penetration depth than the previous case. It has been observed that the jets decay faster with an increasing wall mode wavenumber, likely following the wall mode penetration depth predictions \cite{herrmann1993asymptotic}. Moreover, the scaling relation for wall mode precession frequency was claimed to connect the quadrupolar vortex and BZF observed in the turbulent regime. The behavior regimes of wall modes and internal flow in a moderate $Pr$ RC have been further discussed in more detail in Zhang et al. 2024 \cite{zhang2024wall}. 

These studies provide important insights to liquid metal RC behaviors in a low aspect ratio cylinder. However, the observation of liquid metals RC has been interleaved. We hypothesize that due to the high conductivity and low viscosity nature of liquid metals, the heat and momentum transport of wall modes could be more dominant at low supercriticality, leading to a more prominent quadrupolar vortex structure.

\subsection{Stationary geostrophic mode}

When thermal inertia overcomes the rotation constraint, the liquid metal RC can develop quasisteady modes in columnar shapes aligned with the rotation. Asymptotic theory in the limit of small $Ek$ with a horizontally infinite plane layer and subject to isothermal boundaries give critical values independent from $Pr$ \cite[e.g.,][]{chandrasekhar1961hydrodynamic,julien1998strongly},
\begin{equation}
    Ra_s^\infty \approx 8.7 Ek^{-4/3},\quad \ell_s^\infty \approx 2.4 E^{1/3}.    
\end{equation}
However, $Ra_s^\infty$ is higher than the $Ra$ number of all cases in this study. For example, $Ra_s^\infty = 1.218\times10^9$ for $Ek = 7.77\times 10^{-7}$, which is still slightly larger than the highest $Ra$ from \textbf{RC2kW}. Therefore, we do not expect to clearly observe this mode in this study.



\section{Methods}
\label{method}
\subsection{Laboratory Experimental Setup}
\begin{figure}[ht!]
    \centering 
    \includegraphics[width=\textwidth]{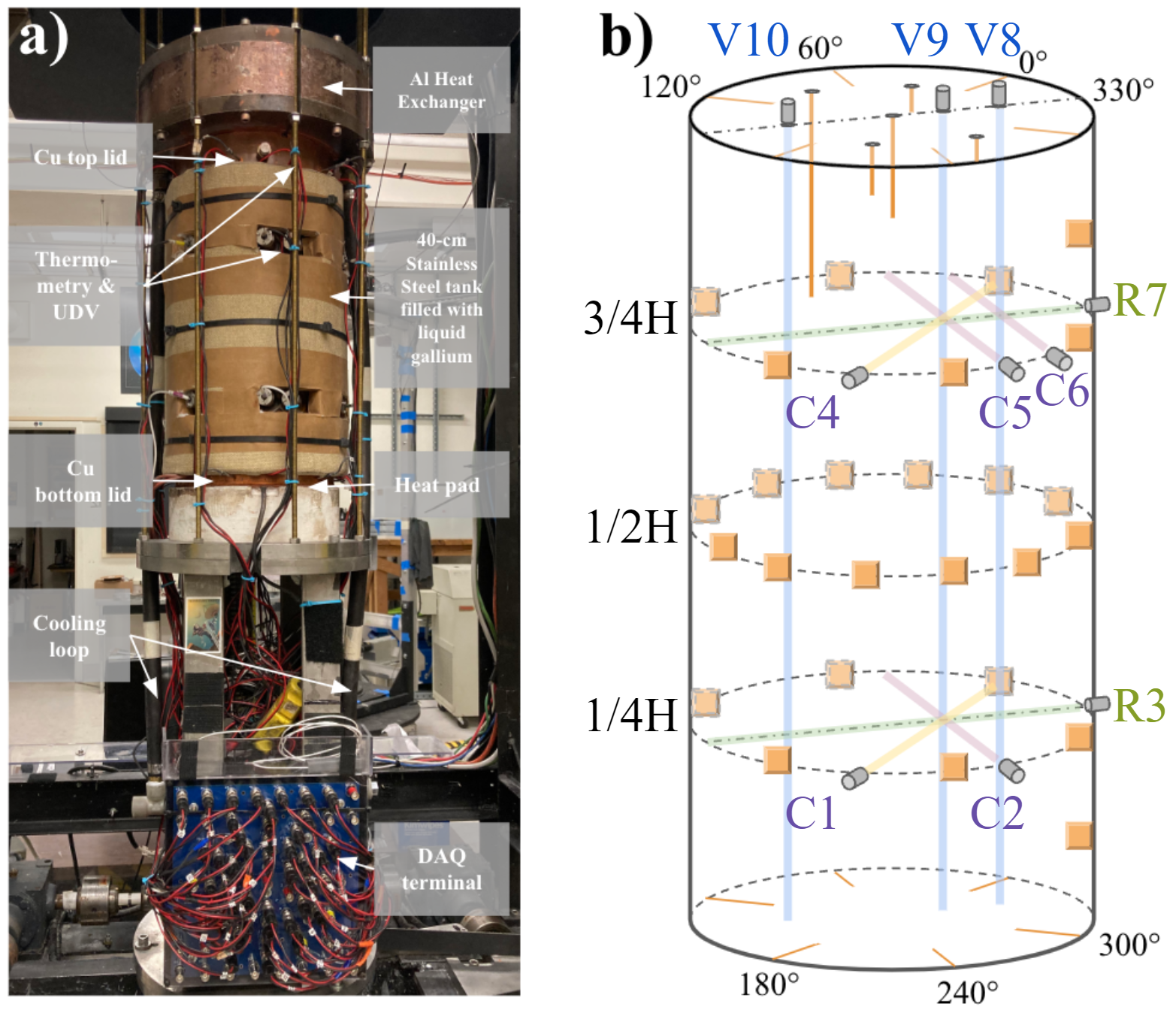}	
    \caption[]{Experimental setup on $\RoMag$ with 40-cm tank. a) A photo of the apparatus. b) DAQ transducers and their lines of sight; internal and sidewall thermometry (orange squares) on the cylinder tank.}
    \label{fig:40setup}%
\end{figure}
%
The laboratory experiments for this study were conducted using UCLA's \RoMag\, device. The cases are named after raw input power at the bottom heat pad. For instance, \textbf{RC60W} and \textbf{RC2kW} have a heat pad power setting around $60 \mathrm W$ and $2 \mathrm{kW}$, respectively. For all the cases, the Reynolds number ($Re = U H/\nu $) is estimated with the maximum velocity measured in our data, $5\times10^2 \lesssim Re \lesssim 5\times10^4$, which is commonly considered turbulent \cite{davidson_2016}. Detailed parameters for these four cases are listed in \cref{tab:para}.

\Cref{fig:40setup}a) shows a photo of the apparatus with descriptions of its components. The container at the center is filled with liquid gallium ($Pr = 0.027$). The end blocks of the container were made of copper. The sidewall is made of stainless steel with a thickness of $3.175\, \mathrm{mm}$. and has a height of $40$ cm so that the aspect ratio of the container is $\Gamma = 1/2$. The container sits on top of a rotating pedestal which is connected to a slip ring and a rotary union for electrical and coolant supply, respectively. The heat exchanger regulates the top surface temperature of the container. An electric heat pad supplies a constant heat flux from the bottom. The sidewall is further insulated by a 2-cm layer of Aerogel. 

A total of 43 thermistors were installed on the container, illustrated with yellow lines and squares in \Cref{fig:40setup}b). Twelve high-precision stainless steel tip thermistors were placed inside the top and bottom boundary at $28.9$ mm radially inwards from the side. The end-block thermistors are used to measure the Nusselt number, which characterizes the ratio of total heat transfer over conduction\cite{ecke2023turbulent},   
\begin{equation}
    Nu = \frac{qH}{k \Delta T}. 
\end{equation}
where heat flux is $q = 4P/(\pi D^2)$, $P$ is the effective heating power with the heat loss correction, $k$ is the thermal conductivity of liquid gallium \cite{aurnou2018rotating}, and $\Delta T$ is measured by the averaged temperature difference between six top and six bottom thermistors. These lid thermistors are marked as short orange lines on the top and bottom of the tank in \Cref{fig:40setup}b). Five high-precision internal thermistors with various tip lengths probe into different depths of the bulk fluids from the top lid. The thermometry has a $10\, \mathrm{Hz}$ sampling rate.

Thermal information near the sidewall can be acquired from a 2D array of temperature measurements circumfering the sidewall exterior because of the high thermal conductivity of the sidewall and the fluid \cite{xu2022thermoelectric,grannan2022experimental,xu2023transition}. We effectively used the sidewall thermistors as velocimetry of the fluid near the sidewall. Twenty-six sidewall thermistors were arranged as follows: six at one-quarter height with $60^\circ$ apart, twelve at the midplane with $30^\circ$ apart, and six at three-quarter height with $60^\circ$ apart. Two additional thermistors were located at $1/8$ and $7/8$ heights, composing a vertical array of five at $330^\circ$ azimuth direction. A few more thermistors not shown in the schematics were also installed to measure sidewall heat loss, room temperature, and expansion tank temperature. 

A total of ten Ultrasonic Doppler probes, shown in \Cref{fig:40setup}, were implemented to provide one-dimensional line-of-sight velocity profiles. They were employed to acquire data with the thermistors at the same time. The chord and radial probe holders are visible on the sidewall as shown in \Cref{fig:40setup}a). \Cref{fig:40setup}b) shows the schematics of the setup: two chord probes, C1, C2, and one radial probe R3 were located at $1/4 H$; three chord probes, C4, C5, C6, and one radial probe R7 were at $3/4 H$. Moreover, three vertical probes, V8, V9, and V10, were placed at the top lid. The probe pairs (C1, C4), (C2, C5), and (R3, R7) are each aligned vertically. The vertical probes are positioned along $330^\circ$ azimuth to intersect with chord and radial probes on the horizontal plane at both heights. The simultaneous measurements of velocity and temperature allow us to observe interacting dynamics at the sidewall and bulk and provide direct evidence of the system's heat and momentum transfer, both locally and globally. 

The Ultrasonic Doppler Velocimetry (UDV) measurements were obtained via a DOP3010 from Signal Processing SA. 
%
%
For our experimental cases, the UDV had various spatial resolutions close to $\approx 2$ mm and temporal resolutions $\approx 0.1$ sec. 

\begin{table}[ht]
\footnotesize
\centering
\begin{tabular}{lccccccc} 
\hline \hline \\[-7pt]
Case    & $\Omega \mathrm{(RPM)}$   & $Ra$    & $Nu$       & $Ek$         & $Ro_\mathcal{C}$        &$Ra/Ra_o$ &$Ra/Ra_w$  \\[2pt]
\hline \\[-7pt]
RC12W (RH)   &  20.154   & 5.79E7    & 1.40      & 5.03E-7     & 0.024    & 0.78    & 0.95   \\[2pt]
RC24W (RH)   &  20.693   & 1.10E8    & 1.55      & 5.09E-7     & 0.032    & 1.42    & 1.73   \\[2pt]
RC60W (LH)   &  12.165   & 1.52E8    & 2.75      & 8.62E-7     & 0.065    & 3.41    & 3.61   \\[2pt]
RC2kW (LH)   &  11.946   & 1.01E9    & 15.83     & 7.77E-7     & 0.160    & 25.91   & 24.43  \\[2pt]
\hline \hline
\end{tabular}
\caption[]{Control parameters and onset frequencies of four liquid gallium rotating convection experiments in a 40-cm cylindrical tank. \textbf{RC12W} and \textbf{RC24W} are subject to a right-hand (RH) rotation, whereas \textbf{RC60W} and \textbf{RC2kW} undergo left-hand (LH) rotation. }.
\label{tab:para}
\end{table}
%


\subsection{Analytical tools}
Vortical structures aligned with the rotation axis are fundamental system-scale features and are responsible for heat and mass transport in rotating convection \cite[e.g.,][]{aubert2001systematic, eckert2002velocity, sprague2006numerical, gillet2007experimental, king2012thermal, vogt2013spin, cheng2015laboratory}. While 1D UDV cannot directly measure the entire vortical structure, we can compare UDV measurements with modeled vortices to interpret and characterize their general morphology and time-dependent dynamics. 

There are several methods to generate synthetic vortices. One approach maps potential flow point vortices onto a disk, resulting in singularities at the vortex centers. Another uses the disk's normal modes, which maintain finite velocity at these centers. Each method has its appeal: potential flow vortices are conceptually simpler, with vorticity only at the centers, while the normal mode method allows the experimental flow field to be decomposed onto these modes at any time.

A group of potential vortices with $m$-fold rotational symmetry have velocity potentials \cite{milne-thomson_theoretical_1938},
\begin{equation}
    \Phi (z, a, \mathcal C) = \sum_{j=0}^{2 m -1} (-1)^{j+1} i \frac{\mathcal C}{2\pi} \log (z + a e^{j- i \pi/m}),
\end{equation}
where $m$ is the azimuthal wavenumber, $z = x+iy$ is the complex coordinates on a plane, $a$ is the half distance between vortex centers, and we assume all vortices have the same circulation $\mathcal C$. Use the Milne - Thomson circle theorem to place a circular boundary of an arbitrary radius of 1 around the vortices,
\begin{equation}
    \Phi_c (z, a, \mathcal C) = \Phi (z, a, \mathcal C) + \bar \Phi \left(\frac{1}{\bar z}, a, \mathcal C \right),
\end{equation}
where $\bar \Phi$ and $\bar z$ are the complex conjugate of $\Phi$ and $z$, respectively, so that $\bar z = x-iy$. The 2-D velocity field $\boldsymbol{u} (u_x, u_y)$ can be written as
\begin{equation}
    u_x = Re \left[ \frac{\partial \Phi_c (z e^{-i \omega t}, a, \mathcal C)}{\partial z} \right], \quad u_y = - Im \left[\frac{\partial \Phi_c (z e^{-i \omega t}, a, \mathcal C)}{\partial z} \right],
\end{equation}
and $\omega$ is the vortex angular frequency. 

For simplicity, we chose point vortices to construct models for the flow structures and generated synthetic Doppler lines to compare with data. Details of normal modes methods can be found in the appendix.

\subsection{Autocorrelation and Pearson's correlation coefficient}

An important goal in our experiments is to determine the characteristics of the turbulent flow. Several methods can be implemented to acquire information on the state of turbulence. Most importantly, wavelength information is required to be extracted from the experimental data. 

The flow's spatial structure and characteristic length scale can be evaluated using the spatial-temporal autocorrelation method on the velocity data acquired by the UDV. Spatial-temporal autocorrelation has been proven successful in numerical simulation and particle image velocimetry (PIV) \cite[e.g.,][]{nieves2014statistical,rajaei2017exploring,madonia2021velocimetry}. For instance, the 1-D spatial autocorrelation between a scalar variable $f$ at position $x$ and $f$ at ($x+\xi$) gives \cite{madonia2021velocimetry},
\begin{equation}
        R_{f}(\xi) = \frac{\left<f(x) f(x + \xi) \right>}{ \left<f^2(x)\right>},
    \label{eq:autocorr}
\end{equation}
where $x$ is the 1-D position vector of the measurements along the ultrasound beam, and $\xi$ is the spatial correlation lag connecting any two points along the measurements. The brackets denote spatial averaging of all the variables. The spatial autocorrelation $R_{f}(h)$ can be calculated by the Fourier convolution theorem, and the same technique can be implemented in time to find the temporal autocorrelation \cite{nieves2014statistical}. 

%

Radial probes R3 and R7, chord probes C2 and C5 are two pairs of Doppler transducers located at the same azimuthal angle but at different heights ($1/4H$ and $3/4H$). To characterize the vertical variation of the flow structure, we calculate Pearson's correlation coefficient (PCC) of velocity data between two probes in each pair, respectively. For instance, given a pair of data set from R3 ($U_i$) and R7 ($V_i$), $\left\{ (U_1, V_1), ..., (U_N, V_N) \right\}$, PCC is represented by $r_{p}$, 
\begin{equation}
    r_{p} (U,V,t)=\frac{\sum_{i=1}^N\left(U_i-\left<U\right>\right)\left(V_i-\left<V\right>\right)}{\sqrt{\sum_{i=1}^N\left(U_i-\left<U\right>\right)^2} \sqrt{\sum_{i=1}^N\left(V_i-\left<V\right>\right)^2}},
\end{equation}
where $i$ is the sample index of velocity measurements along the line, $N$ is the sample size, and $\left< ...\right>$ is the sample mean in 1D line space. PCC provides a quantitative measure of the correlation between velocity fields: if $r_p \gtrsim 0.5$ for the given pairs listed above, the velocity of two different heights are correlated, which implies the flow structure is approximately axially invariant.


%
%

\section{Precessing wall mode is the dominant thermal feature}
\label{wallmode}
\begin{table}[ht]
\centering
\footnotesize
\begin{tabular}{lcccccccc} 
\hline \hline \\[-7pt]
Case    & $\widetilde{f_o}$ & $\widetilde{f_w}$  & $\widetilde f(T_{int,c})$ & $\widetilde f(T_{int,2})$ & $\widetilde f(T_{sw})$ & $\widetilde f(u_z)$   & $\widetilde f_2(u_z)$    & $\widetilde f(u_c)$  \\[2pt]
\hline \\[-7pt]
RC12W    & 0.1014     & 0.0022    & 0.0026    & 0.0026      & 0.0026   & 0.0027    & 0.0356   & 0.0027 \\[2pt]
RC24W    & 0.1014     & 0.0022    & 0.0058    & 0.0058      & 0.0058   & 0.0059    & 0.1369   & 0.0059 \\[2pt]
RC60W    & 0.1148     & 0.0038    & -         & 0.0190      & 0.0190   & 0.0202    & 0.2129   & 0.0404 \\[2pt]
RC2kW    & 0.1156     & 0.0038    & -         & 0.0555      & 0.0496   & 0.0522    & -        & 0.0500 \\[2pt]
\hline \hline
\end{tabular}
\caption[]{Onset frequencies and measured peak frequencies from thermometry and velocimetry. The dimensional onset frequency solution for sidewall ($f_w$) and oscillatory bulk modes ($f_o$) are calculated from the linear analysis with semi infinite wall (Zhang \& Liao). We normalized these values by the rotation frequency of the system $f_\Omega = \Omega/(2\pi)$, so that $\widetilde{f_o} = f_o/f_\Omega$, and  $\widetilde{f_w} = f_w/f_\Omega$. The peak frequency is marked by "-" when no significant peak can be detected from the spectra.}
\label{tab:fft}
\end{table}
The sidewall and internal thermometry were used to characterize the wall modes and oscillatory modes in liquid metal rotating convection. UDV has also been used to probe the interior velocity profile and find correlations with the thermometry. 
\begin{figure}[ht!]
	\centering 
	\includegraphics[width=\textwidth]{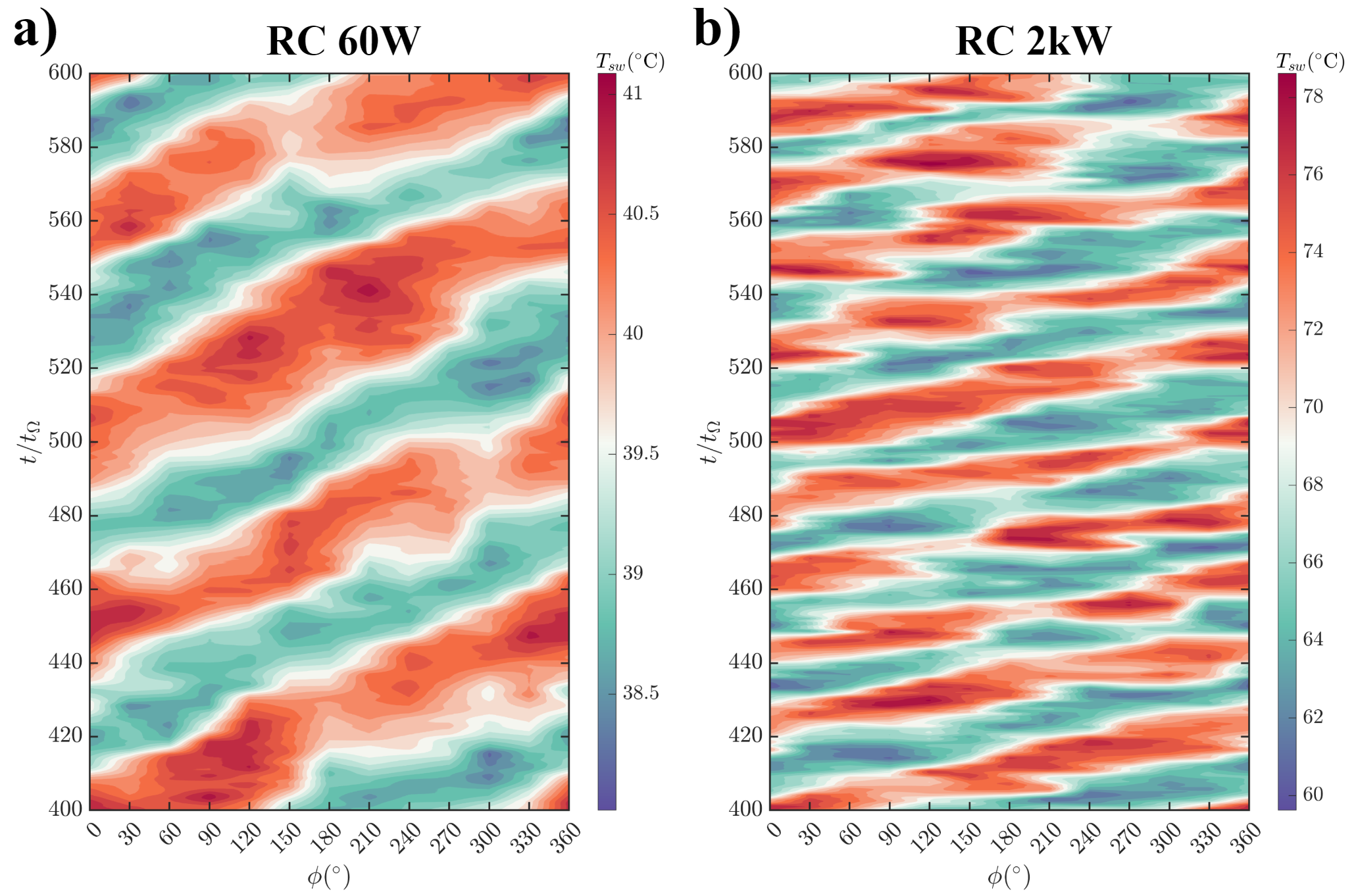}	
	\caption[]{Hovm\"oller diagrams of sidewall midplane temperature contour over time for a) the \textbf{RC60W} and b) \textbf{RC2kW} case. The contour plots were interpolated by thermometry data every $30^\circ$ in the azimuthal direction. These case were taken over $1.6\tau_\kappa$, where $\tau_\kappa = H^2/\kappa$ is the thermal diffusion time and is equivalent to more than $4\times 10^3 t_\Omega$. Only $200\, t_{\Omega}$ data are shown here for comparison.     
}\label{fig:sw_hov}%
\end{figure}
%

\Cref{fig:sw_hov} shows a pair of azimuthally oriented spatial-temporal temperature color maps of the sidewall midplane for \textbf{RC60W} and \textbf{RC2kW}. The red-orange color represents higher temperature, while the blue-green color represents lower temperature with respective to the mean temperature of each case. As time evolves, liquid metal carries the local thermal anomalies from a lower azimuthal angle to a higher one in a right-hand direction, which means that the local flow near the sidewall is processing in a retrograde direction of the rotation (left-handed). Moreover, only one hot and one cold structure existed along the entire azimuth for both cases at each time step, indicating this is a mode number $m=1$ wall mode. Another simple observation is that the wall mode in \textbf{RC2kW} has a higher frequency than that of \textbf{RC60W}, as both cases are shown in the same timescale of $200\, t_\Omega$. This is further investigated in the spectral analysis below.  

We performed spectral analysis on multiple thermometers located at sidewall and internal at two different depths and radius locations. One is located at the center at $55$ mm deep from the top fluid layer, another is near $2/3R$ at $105$ mm deep from the top fluid layer. Moreover, spectra of vertical and chord velocities are also included for comparison. The peak frequencies from the spectra are summarized in \Cref{tab:fft}.  
\begin{figure}[ht!]
	\centering 
	\includegraphics[width=\textwidth]{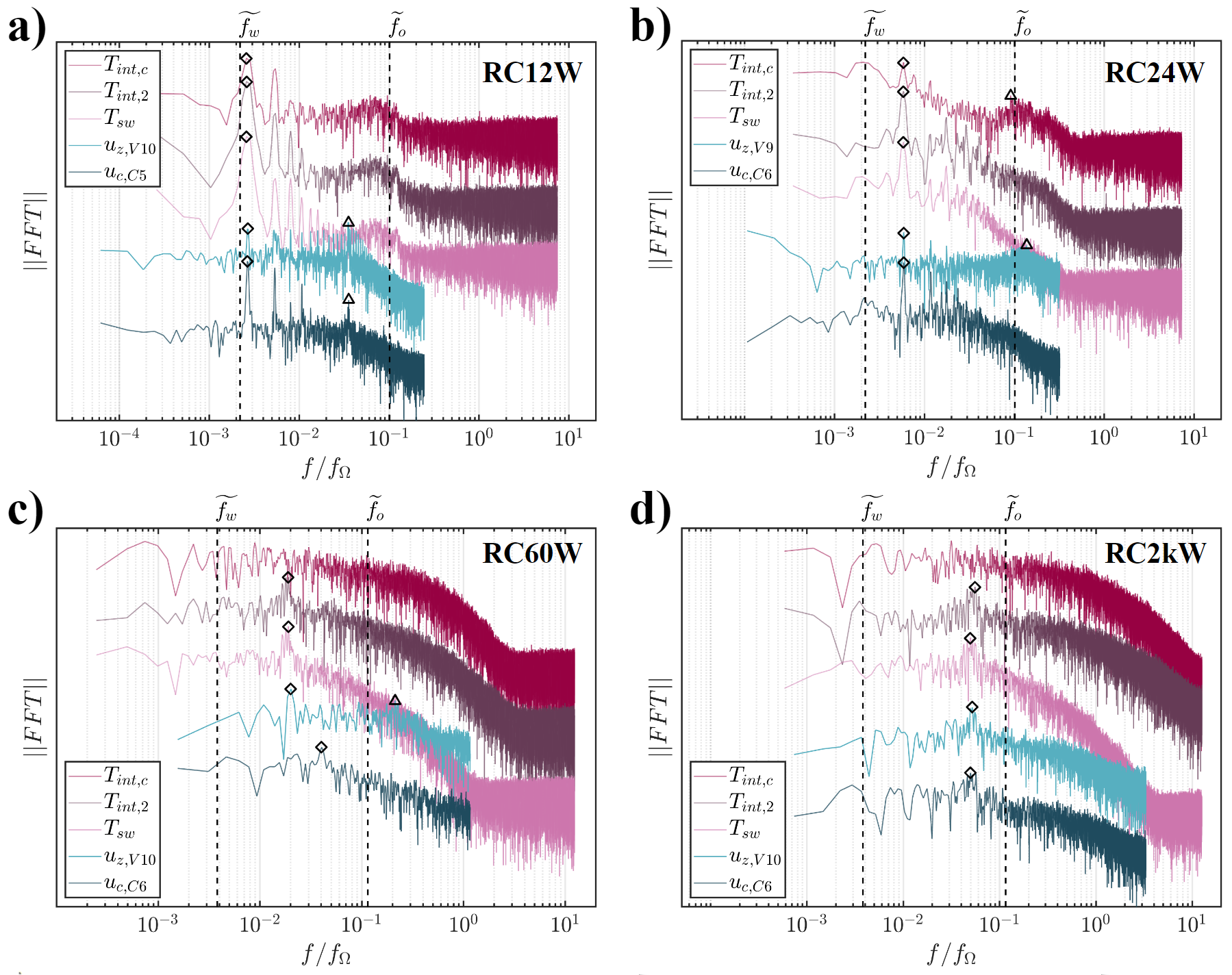}	
	\caption[]{Thermovelocimetric spectra of a) \textbf{RC12W}, $Ro_C = 0.024$. b) \textbf{RC24W}, $Ro_C = 0.032$, c) \textbf{RC60W}, $Ro_C = 0.065$, d) \textbf{RC2kW}, $Ro_C = 0.160$. The spectra of the center interior thermistor data are labeled as $T_{int,c}$; the other interior thermistor's data are labeled as $T_{int,2}$; the sidewall thermistor located at $7/8 H$ is marked as $T_{sw}$; $u_{z}$ is the spectra of vertical velocity data from V10 or V9; and $u_c$ is the spectra of chord velocity data from C5 or C6. The FFT powers are arbitrarily set to avoid overlapping and better display of the spectra. The vertical black dashed lines mark wall mode onset frequency $\widetilde{f_w}$, and oscillatory mode frequency $\widetilde{f_o}$ for each case, and normalized by the rotation frequency, $f_\Omega$. The diamond symbols ($\diamond$) mark the primary peak frequency related to wall mode, while the triangle symbols ($\triangle$) label the peak frequency relevant to oscillatory mode.   
}\label{fig:TV_FFT}%
\end{figure}
\Cref{fig:TV_FFT}a) shows that wall mode in \textbf{RC12W} is observable in all diagnostics listed in the figure, and the primary spectra peak frequencies match the onset prediction for wall modes. Unknown ripples contribute to a secondary peak in velocity data, see \ref{app1} for velocity profile. 

\Cref{fig:TV_FFT}b) shows that the wall mode frequency in \textbf{RC24W} is still dominant and was picked up by all the probes. However, the peaks shifted to a higher frequency. Internal thermistor and vertical velocity have a secondary peak near the predicted oscillatory mode frequency, while the chord probe did not detect this mode. This is expected as the oscillation is the most dominant in the vertical direction.    

\Cref{fig:TV_FFT}c) shows that in \textbf{RC60W}, wall mode continues to grow faster, chord peak is double the wall mode, indicating a steady $m=2$ mode located internally. Further analysis is presented with the following velocity profiles. The vertical velocity spectra show the oscillatory mode grows to higher frequencies than in \textbf{RC24W}. As supercriticality increases, the growing peak of the oscillatory mode has also been observed from previous liquid metal experiments with a similar setup \cite{vogt2021oscillatory}. Moreover, the center thermistor did not detect the wall mode as bulk turbulence increased.

\Cref{fig:TV_FFT}d) shows that the center thermistor in \textbf{RC2k} displayed broadband turbulence with no peak. All the rest of the probes detected wall mode at the highest supercriticality of $Ra/Ra_w = 24.43$ among all the cases in this study, demonstrating the continuation of growth in wall mode frequency with an increasing supercriticality.  

\begin{figure}[ht!]
	\centering 
	\includegraphics[width=\textwidth]{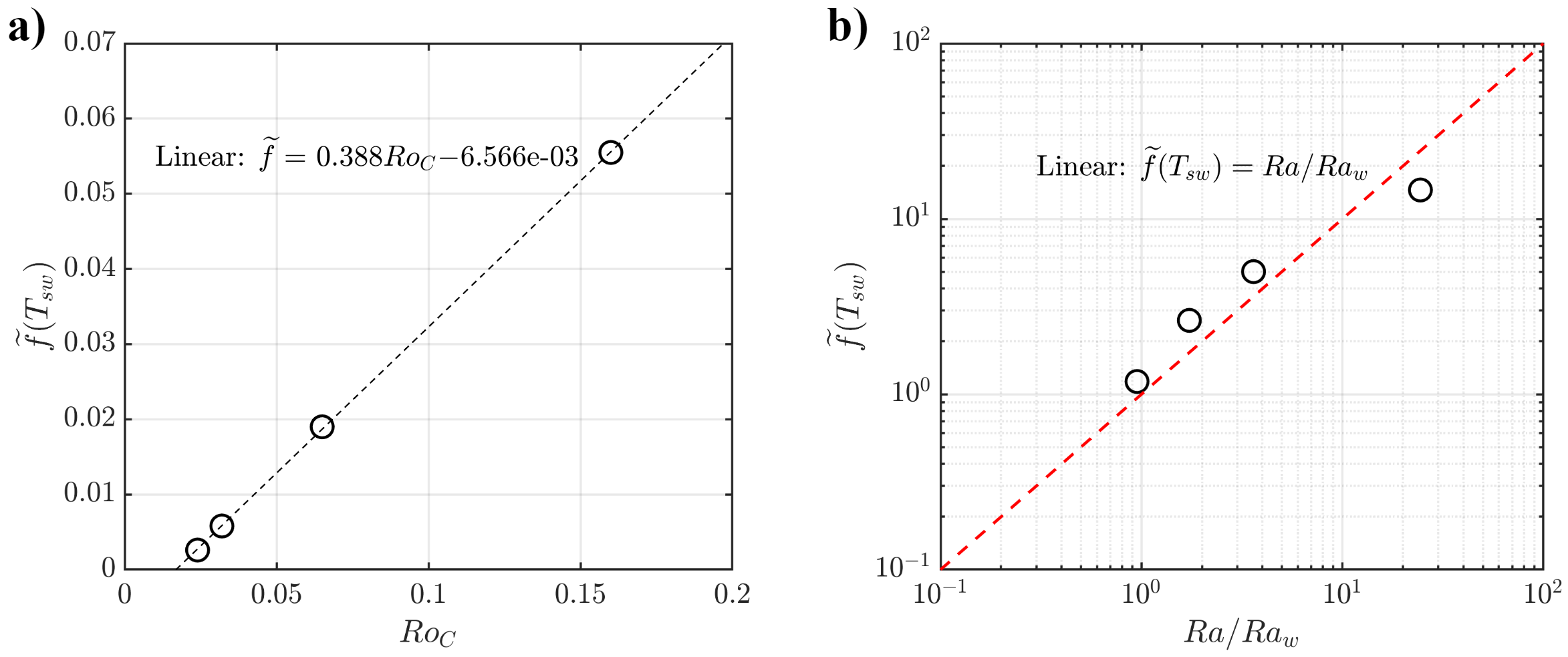}	
	\caption[]{Sidewall peak temperature frequencies of the four cases as a function of a) convective Rossby number $Ro_C$, b) wall mode supercriticality $Ra/Ra_w$.
}\label{fig:fwRoc}%
\end{figure}

\Cref{fig:fwRoc} shows wall mode frequency from the sidewall thermistor data as a function of convective Rossby number $Ro_C$ and supercriticality $Ra/Ra_w$. We find a linear empirical law between the wall mode frequency above onset with $Ro_C$, 
\begin{equation}
    \widetilde f(T_{sw}) = \omega_w/\Omega = 0.388 Ro_C - 6.566\times 10^{-3},
    \label{eq:fw_Roc}
\end{equation}
where $\omega_{sw}$ is the wall mode drift angular speed. These four frequency data also approximately agree with the linear scaling of wall mode supercriticality, $\widetilde f(T_{sw}) \approx Ra/Ra_w$. However, with limited data, we cannot confirm which is a better fit for our data between these two scalings. 

\section{Quadruplar cyclone revealed from velocity profiling}
\label{quad}

\begin{figure}[ht!]
	\centering 
	\includegraphics[width=\textwidth]{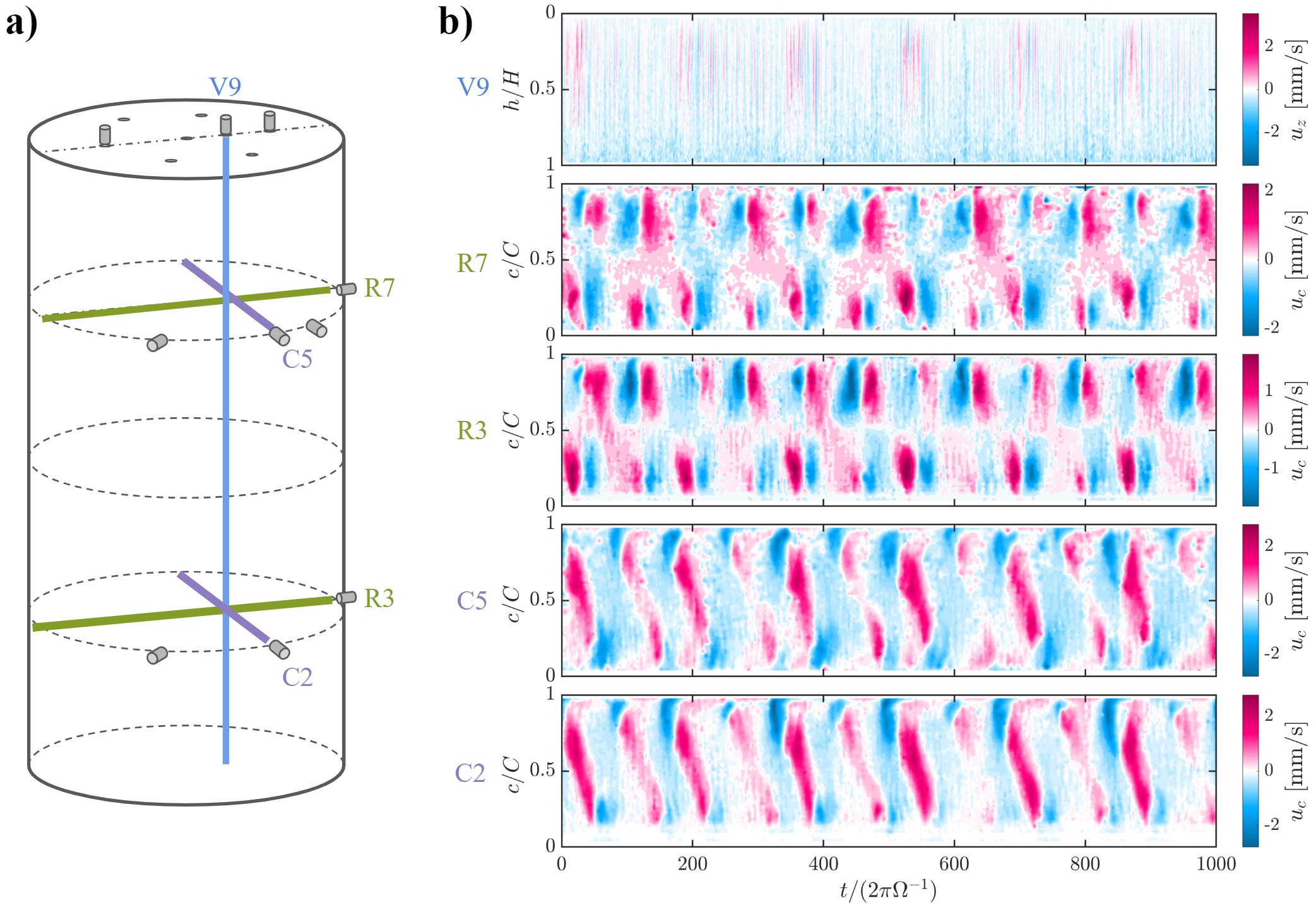}	
	\caption[]{UDV profiles of \textbf{RC24W} subject to a right-hand rotation. The vertical probe V9's velocity data show periodic upwelling/downwellings due to the presence of the wall mode. There are strong correlations between R7 and R3, C5 and C2, respectively. Note that the color stripe in C2, C5, and D6 are have opposite slope from chord profile in \textbf{RC60W}, because these two cases have the opposite rotation direction. The asymmetry of color in each chord/radial data could possibly be due to the difference in the radii of the cyclones. The wall mode remains a retrograde feature in all the cases.}
\label{fig:rc24w_dop}%
\end{figure}
The thermometer array at the sidewall makes it challenging to detect bulk dynamics in the slender tank. There are a limited number of internal thermistors, and they are located near the top boundary within a quarter tank height. More importantly, thermal signals of the wall modes are dominant throughout the radius (i.e., \Cref{fig:TV_FFT}), inhibiting our ability to probe into the fluctuations and flow velocity in the bulk. Therefore, UDV is required to investigate this internal momentum transfer of liquid metal. 


\Cref{fig:rc24w_dop} are velocity Hovm\"oller diagrams of the \textbf{RC24W} case at five distinct paths into the bulk of the liquid gallium, as shown on the left schematics. In this case, $Ra = 1.10\times 10^8$, $Ek = 5.09\times 10^{-7}$, and $Ro_\mathcal C = 0.032$ (see \Cref{tab:para}). The contour plots in \Cref{fig:rc24w_dop} show an equilibrated 1-D velocity timeseries at five probe locations over a time interval of $9363 \tau_\Omega$. The vertical axes for all five panels are distances normalized by the total path length of the Ultrasonic beam. The red color in vertical probe V9 data represents positive velocity in upwelling flows, and blue represents downwelling flows. For chord probes (C5, C2) and radial probes (R7, R3), red means fluid moving away from the probe, whereas blue means fluid moving towards the probe.  

The comparison between the vertical (V9) and horizontal velocity data (R7, R3, C5, \& C2) highlights the anisotropy in the velocity field of \textbf{RC24W}. The maximum velocities in the horizontal planes, measured by chord and radial probes, are larger than the vertical velocity in V9. Additionally, the velocity Hovm\"oller diagram of V9 shows a significantly different characteristic frequency comparing the other horizontal velocities. 

\begin{figure}[ht!]
	\centering 
	\includegraphics[width=0.7\textwidth]{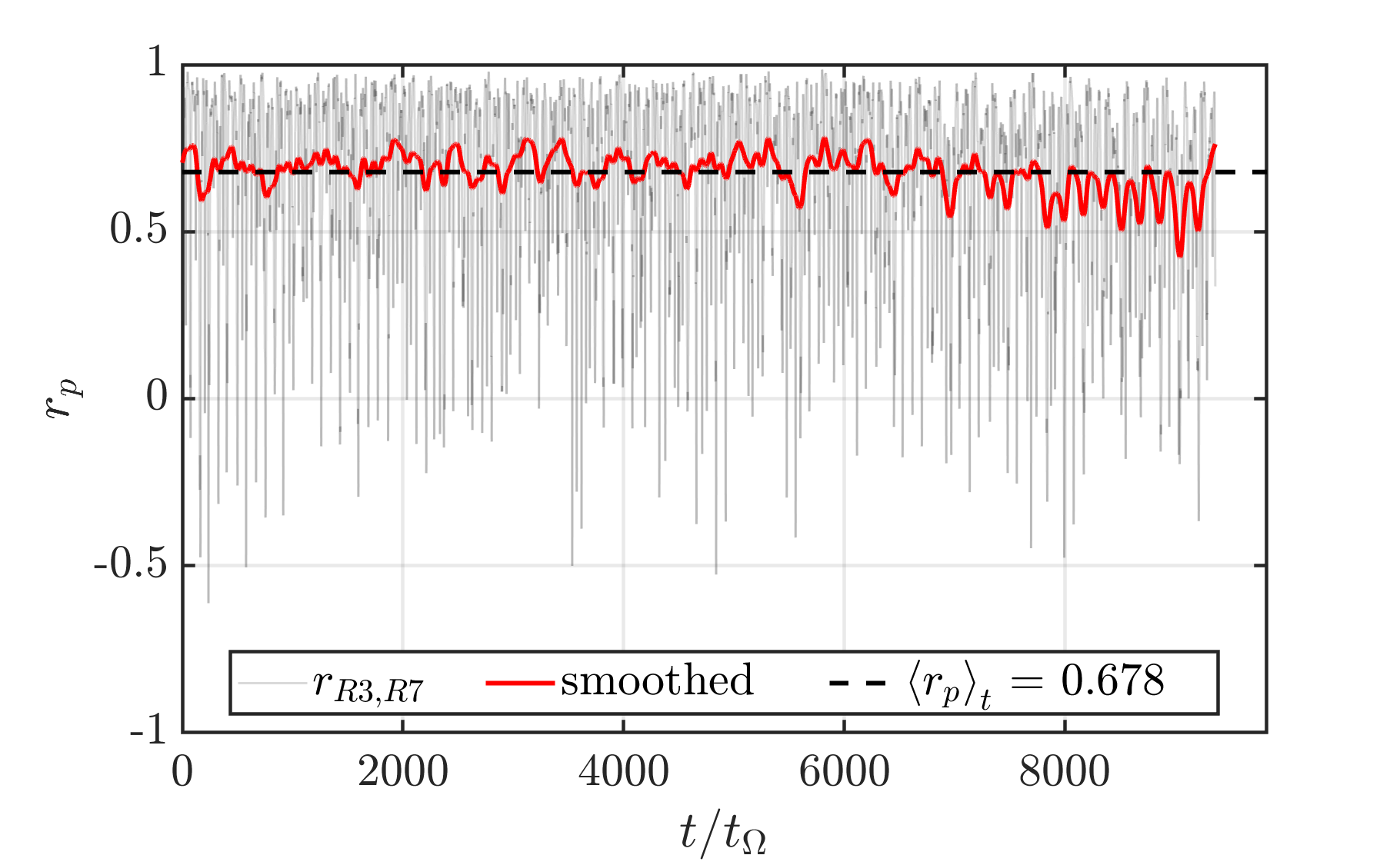}	
	\caption[]{Timeseries of Pearson's correlation coefficient (PCC), $r_p$, between R3 and R7 in \textbf{RC24W}. The grey curve shows raw data points representing correlation values along the diameter paths of R3 and R7 at each timestep. The red curve is the Gaussian smoothed line from the raw data, and the black dashed line shows the time-averaged PCC, $\left< r_p\right>_t$. The time has been normalized by rotation period, $t_\Omega$. 
}\label{fig:pcc24}%
\end{figure}

The bulk flow is quasi-2D, nearly invariant in $\hat{z}$ at $Ro_C = 0.032$. The schematics in \Cref{fig:rc24w_dop} show that the radial probe pair (R7, R3) and the chord probe pair (C5, C2) are located at the same azimuthal angle, respectively, only at different heights of $3/4H$ and $1/4H$. However, the velocity fields measured by R7 and R3 are almost identical to each other, as are C5 and C2. \Cref{fig:pcc24} shows Pearson's correlation coefficient (PCC) between R3 and R7 data as a function of time over the entire case. The time-averaged coefficient is $0.678$, indicating a good correlation between R3 and R7 velocity data. As $Ro_C$ becomes larger and supercriticality increases, PCC gradually decreases as turbulence dissociates the flows at different heights. For \textbf{RC60W}, $\left< r_p\right>_t = 0.486$ between R3 and R7, which is still a moderately good correlation. This result suggests that the flow field of \textbf{RC24W} and \textbf{RC60W} approximately follows the Taylor-Proudman theorem that the fluid motion is invariant along the axis of rotation. We can thus make the assumption that the dominant bulk modes for \textbf{RC24W} and \textbf{RC60W} are axially invariant. 

\begin{figure}[ht!]
	\centering 
	\includegraphics[width=0.65\textwidth]{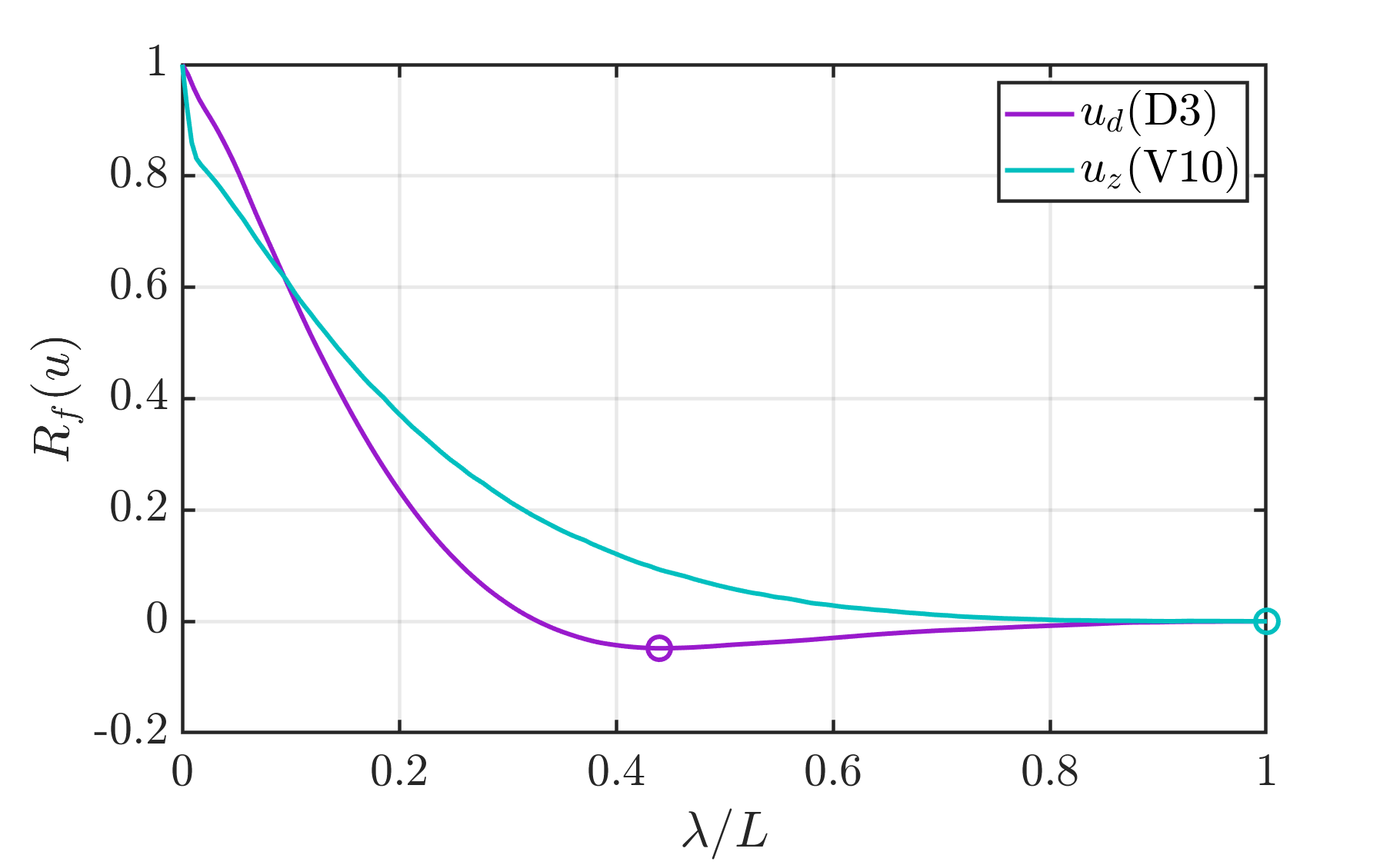}	
	\caption[]{Time-averaged spatial autocorrelation of velocity profile in the radial (R3) and vertical directions (V10) \textbf{RC60W} as a function of characteristic wavelength, $\lambda$. The wavelength is normalized by the total length of the UDV beam, i.e., $L$ is the observable length in the radial direction for R3 and the vertical observable fluid layer height for V10. The minimum point on each curve marks the characteristic spatial periodic wavelength. $R_f<1$ means the spatial pattern is alternative 
}\label{fig:ac60}%
\end{figure}
\begin{figure}[ht!]
	\centering 
	\includegraphics[width=\textwidth]{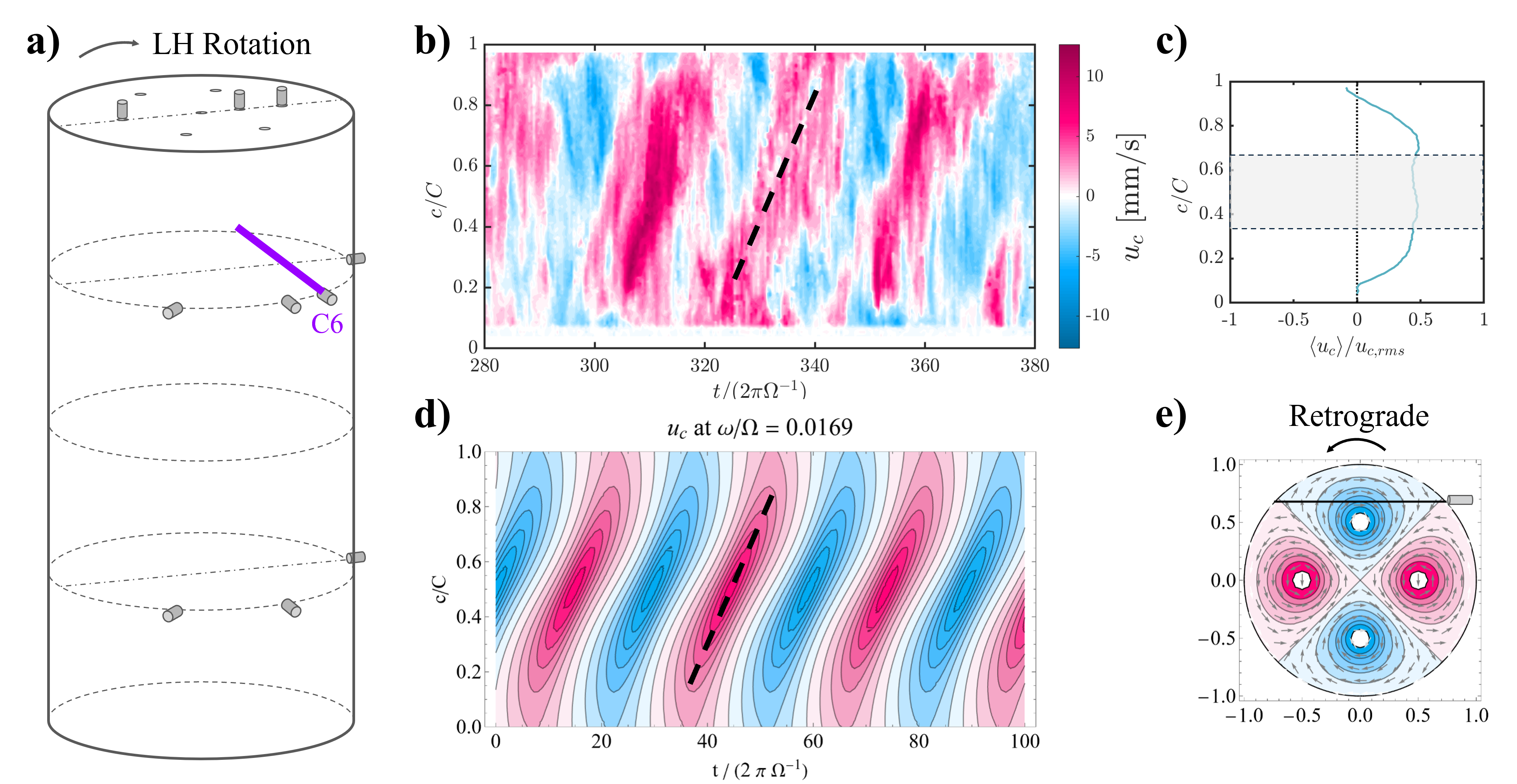}	
	\caption[]{a) Location of the chord probe C6. b) Doppler velocity Hovm\"oller diagram of C6 in RC60W case. c) Time-averaged chord velocity normalized by r.m.s velocity shows a retrograde precession of the structure as the positive direction is in RH, opposite of the system rotation in LH direction. The measured chord velocity can be projected into an approximately constant drifting speed $u_\phi \sim 0.0169 \Omega$, if a solid-body rotation is assumed at the zeroth order. d) Hovm\"oller diagram of the synthetic Doppler C6's data and a horizontal velocity contour of the azimuthal wavenumber $m=2$ synthetic vortices with a constant background retrograde rotation calculated from the mid-chord averaged zonal velocity. The color represents the velocity. e) Top view of the synthetic vortices with retrograde background rotation and C6 Doppler probe position. }
 \label{fig:synth6}%
\end{figure}
\Cref{fig:ac60} shows a time-averaged 1D spatial autocorrelation of velocity in the radial and vertical direction, measured by R3 and V10. The autocorrelation coefficient $R_f$ was computed at each time as a function of wavelength normalized by the total beam length and then averaged to one curve representing the overall spacial characteristic wavelength. The magenta line has a minimum below zero, representing a characteristic wavelength close to $0.5$ of the total diameter, with an anti-phasing periodicity. This means that the velocity field measured along the diameter has two structures with opposite velocities. The cyan curve marks the vertical velocity of the profile and has a minimum at the $\lambda = L$. In this case, the total beam length is approximately the entire height of the tank. Therefore, there is only one characteristic structure present at each timestep. 

\begin{figure}[ht!]
	\centering 
	\includegraphics[width=\textwidth]{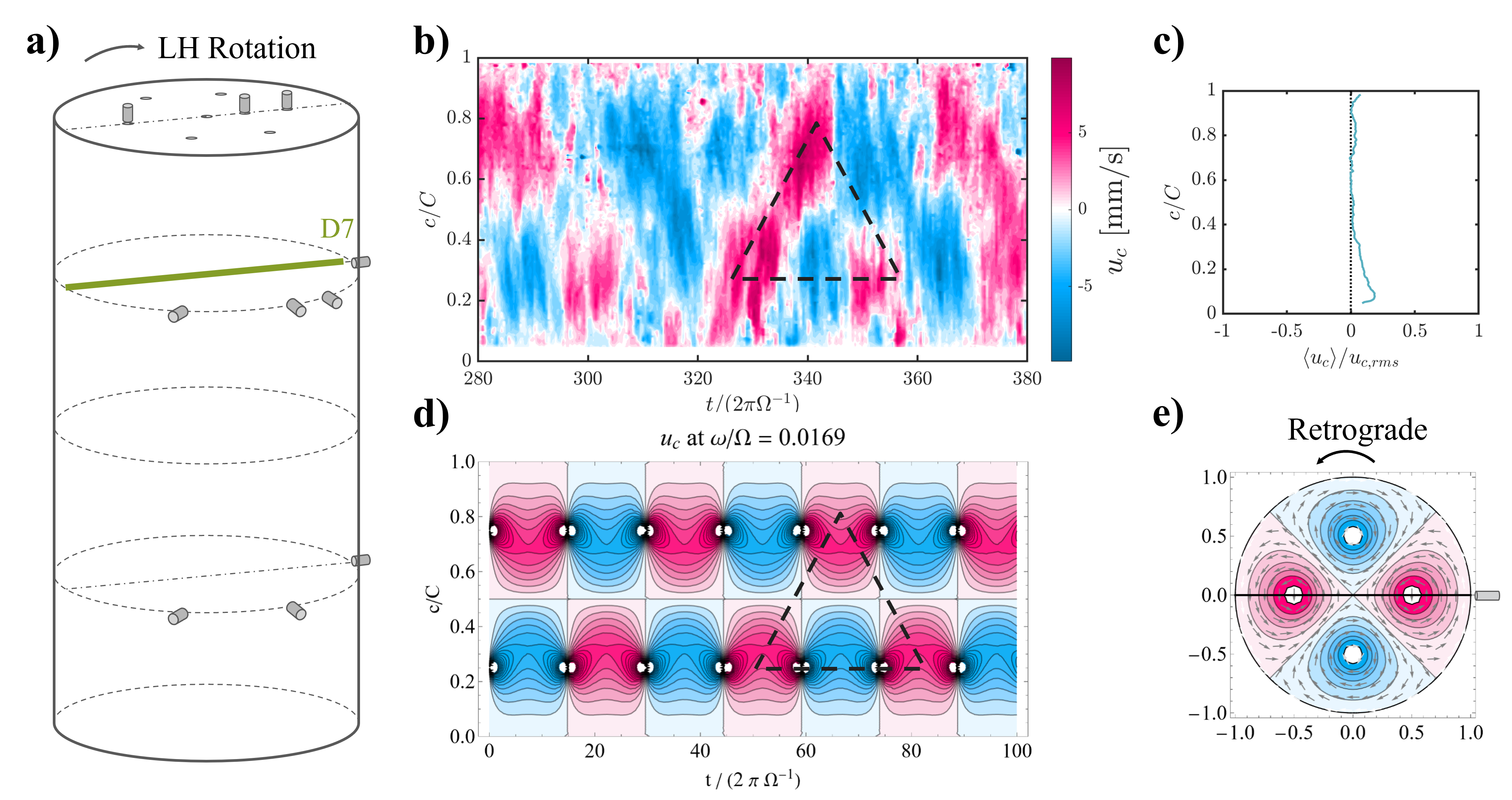}	
	\caption[]{a) Location of the radial probe R7. b) Doppler velocity Hovm\"oller diagram of R7 in \textbf{RC60W} case. c) Time-averaged radial velocity is near zero, indicating the lack of radial evolution of the structure over time. d) Hovm\"oller diagrams of the synthetic Doppler R7's data. The contour map on the right shows the same $m=2$ vortices with a retrograde rotation calculated from chord velocity measured by C6. The color represents the velocity.  e) Top view of the synthetic vortices with retrograde background rotation and R7 Doppler probe position. 
}\label{fig:synth7}%
\end{figure}

Combining the horizontal velocity data taken by chord transducer C6 shown in \Cref{fig:synth6}a) and radial transducer R7 shown in \Cref{fig:synth7}, we hypothesize that a prominent system-scale azimuthal wavenumber $m=2$ quadrupolar cyclone emerges near wall mode onset. \Cref{fig:synth6}b) focuses on a Doppler velocity Hovm\"oller diagram of chord probe C6 in \textbf{RC60W}. We applied a noise filter to remove rare unrealistic signals. The velocity data very close to the boundary had also been removed due to diagnostic limitations. \Cref{fig:synth6}c) shows a time-averaged chord velocity along C6, normalized by the root-mean-square of the measured velocity. The velocity remains mostly constant at the center of the chord. If we assume the time-averaged precession of the bulk is a solid body rotation, the angular velocity of the bulk drift can be inferred from the chord velocity of the probe using the simple relationship below \cite{king2009investigation}:
\begin{equation}
    \boldsymbol{\omega} = \frac{u_c}{r \sin \theta} \hat \phi= \frac{u_c}{R_c} \hat \phi,
\end{equation}
where $\theta$ is the angle between the chord and the line from the point to the center, $r$ is the distance of a point from the center, and $R_c$ is the distance between the center and the chord. 
Taking the average of the middle $30\%$ of the chord velocity (grey area), the net drift of the bulk flow shows a retrograde direction of the rotation of the cylinder at $\omega = 0.0169\Omega$. 

\Cref{fig:synth6}c) is a synthetic Hovm\"oller diagram of the one-dimensional velocity at the same chord location as the C6. This 2-D synthetic velocity field model comprises $m=2$ cyclones with four neighboring opposing point vortices circumscribed by a circle with an imposed net background rotation. The horizontal flow field of this model is depicted in \Cref{fig:synth6}e). The background rotation rate is taken from the time-averaged experimental data. The tilted and alternating velocity patches along the chord are similar to the ones shown in the synthetic velocity Hovm\"oller diagram. The slope of the tilted structure can be explained as a result of a net drift in superposition with two pairs of smaller vortices that each occupies about one radius of the cylinder. 

The $m=2$ cyclones in \textbf{RC60W} also explain the peak frequency in C6's spectrum, as shown in \Cref{fig:TV_FFT}c). The peak frequency $\widetilde f = 0.040$, which is $2.367$ times the drifting frequency. Because the chord probe looks at only one side of the $m=2$ cyclones, it picks up roughly twice the alternating patterns as the cyclone drifts in one period. Interestingly, chord probes in \textbf{RC12W} and \textbf{RC24W} do not have the doubled peak in their spectra. There are two possible explanations. First, wall modes in these near-onset cases are too dominant that the repeating cyclones overlap with the double harmonics of the wall mode peak; secondly, vortices in these cases may have asymmetric radii, contributing to a larger periodicity with one pair of vortices. 

\Cref{fig:synth7}, parallel to \Cref{fig:synth6}, investigates velocity data taken by R7. \Cref{fig:synth7}b) shows a Doppler velocity Hovm\"oller diagram of the radial probe R7 in \textbf{RC60W} case in comparison to a synthetic Hovm\"oller diagram at the exact diameter location as the R7. \Cref{fig:synth7}c) shows that the time-averaged radial velocity is near zero, agreeing with our assumption of a net drift with the near constant angular velocity. The alternating triangular patterns in the experimental data, as shown in \Cref{fig:synth6}d), are captured by the alternating patterns of synthetic data vortices as they pass through the radial beam. The position of the probe is shown in \Cref{fig:synth6}e). The dark spots in the synthetic Hovm\"oller are attributed to artificial errors from the singularity of the synthetic vortex model and thus can be neglected. 

Sufficient evidence suggests that the $m=2$ quadrupolar cyclonic structure is a robust bulk mode that appears at the onset of the wall mode. This structure remains consistent in \textbf{RC12W} and \textbf{RC24W} at lower $Ro_C = 0.0242$ and $0.0323$, just above the wall mode onset (see \ref{app1} for detailed velocity profiles). This further suggests that $m = 2$ cyclone is a dominant stable flow structure at low $Ro_C$, and can describe the velocity field of the bulk flow of \textbf{RC60W} case. However, the structure disappears at a larger $Ro_C = 0.160$.

\begin{figure}[ht!]
	\centering 
	\includegraphics[width=\textwidth]{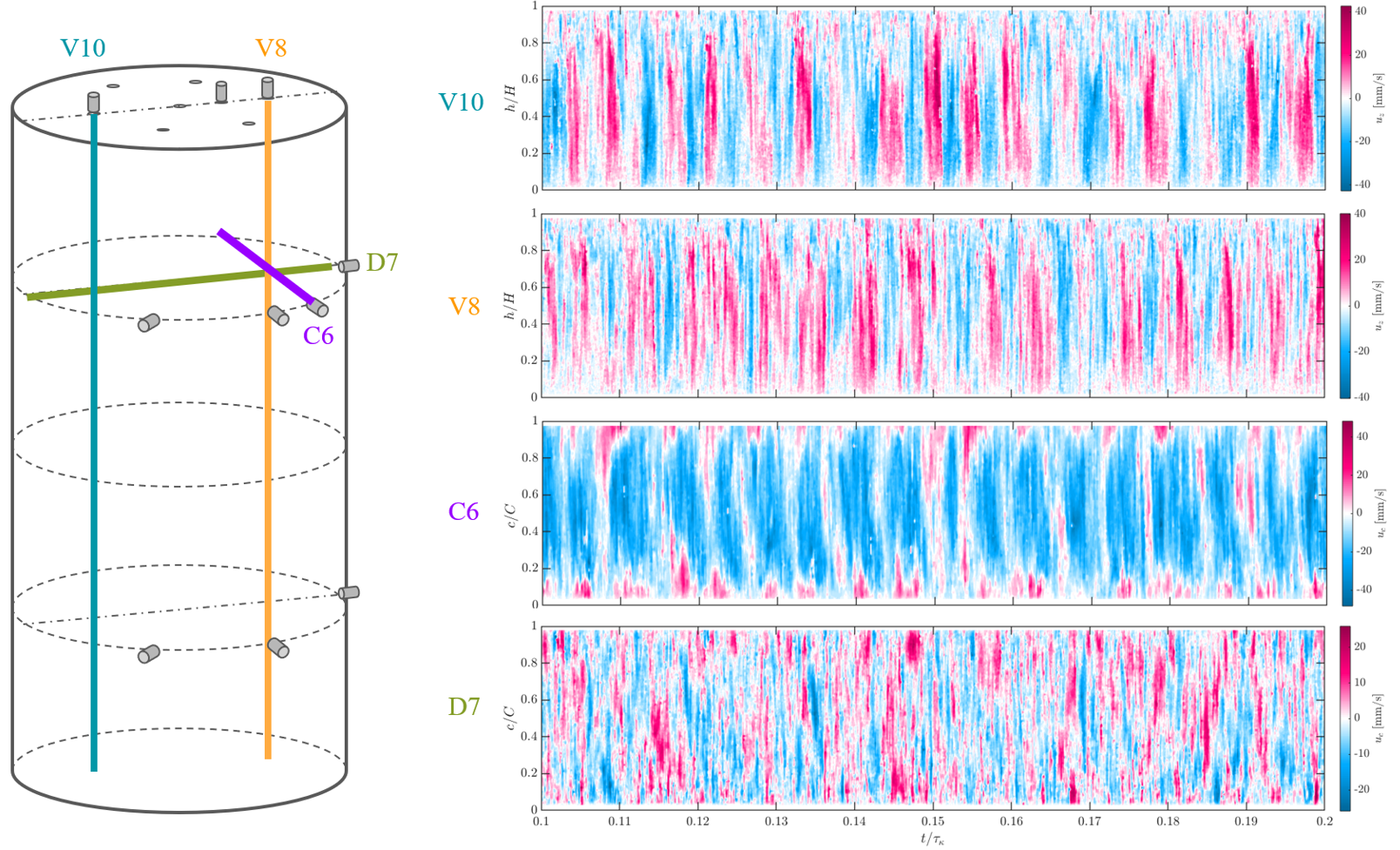}	
	\caption[]{Caption: Hovm\"oller diagrams of UDVs for the \textbf{RC2kW} case. Four Doppler probes are set up as the schematics on the left. Probe R7 is the radial probes, V10 and V8 are the vertical probes, and C6 is the chord probe. The diagrams show the velocity in color along the beam direction. The vertical axes are the distance normalized by the path length. The horizontal axes are time normalized by thermal diffusion time $\tau_\kappa$. All the diagrams show $t= 0.1 \tau_\kappa$ intervals. For vertical probes, red represents positive velocity in the upward direction. For chord and radial probes, red represents velocity moving away.
}\label{fig:dop2kw}%
\end{figure}

\Cref{fig:dop2kw} shows the velocity Hovm\"oller diagrams for four velocity transducers: chord C6, radial R7, vertical V8 and V10 for the \textbf{RC2kW} case. Detailed parameters are displayed in \Cref{tab:para}. In sharp contrast to \textbf{RC60W} case, the structures in D6 and radial R7 are completely different. Strong negative flows in D6 indicate the zonal jet still exists, happening at a much faster speed. No coherent structures can be observed in R7. This is likely because the strong thermal forcing relaxes the Taylor-Proudman constraint. 

No clear correlation or anti-correlation in vertical velocities of V8 and V10 can be detected in \textbf{RC2kW}, $\left< r_p\right>_t = -0.0304$. This agrees with lower supercritical cases, for instance, $\left< r_p\right>_t = -0.032$ for \textbf{RC60W}'s V8 and V10's velocity data. Furthermore, R3 and R7 have $\left< r_p\right>_t = -0.0422$ in \textbf{RC2kW}. Therefore, no correlation can be found for the transducers at the same azimuthal angle, in contrast to the lower supercritical cases. This observation suggests that the flow became three-dimensional in \textbf{RC2kW}, and a quadrupolar cyclonic structure can no longer exist. 

\section{Discussion}
\label{disc}


\subsection{Further validation of the quadrupolar cyclone model}

We further compare the velocity profile of C6 and R7 with $m>2$ modes using the same azimuthal precession speed (see \ref{app3}). Only $m = 2$ vortices with a background zonal flow can explain the bimodel structure in the chord and radial velocity data. At any given moment on velocity measured by C6, the tilted stripes in \Cref{fig:synth6}b) have the same velocity direction spanning the entire chord. With the same net drift, any $m>2$ vortices patterns will result in a mismatch in frequency and wavenumber. Moreover, the alternating triangular shape in radial velocity indicates two alternating structures along the diameter, and the azimuthal wavenumber $m$ has to be even, c.f., \cref{fig:msynth}a). 

The mismatch in patterns and frequency between the experimental data and the synthetic UDV data also rules out the possibility that the internal structures originated from the bulk oscillatory modes. The predicted velocity field is calculated by the normal mode method. More details can be found in \ref{app2}, where \Cref{fig:osc} shows the expected C6 and R7 Doppler velocity at the onset of thermal-inertia oscillatory mode. 

\subsection{Comparison to $Pr \sim 1$ quadrupolar cyclone}

The quadrupolar cyclone in liquid metal RC is qualitatively a much more prominent feature than in $Pr \gtrsim 1$ fluid. The coherent velocity profile of the liquid metal quadrupolar cyclone can be directly measured without time-averaging and has a much larger lateral velocity in the horizontal plane. In $Pr \gtrsim 1$, quadrupolar has a horizontal maximum velocity about $\sim \pm 10\%$ of the vertical velocity \cite{de2023robust}. In liquid metal, the lateral velocity can be even larger than the vertical velocity. 

Robust wall mode has been observed to induce quadrupolar vortex in rotating convection at low $\Gamma$. The mode number of the vortices $m$ depends on the wall mode azimuthal wavenumber, which is ultimately determined by the aspect ratio of the cylindrical container. In our system, the wall mode remains $m=1$. The wallmode drifting frequency grows linearly with $Ro_C$, so that $\omega_{sw}/\Omega \propto EkPr^{-1/2}Ra^{1/2} = Ro_C$ was shown in \Cref{eq:fw_Roc}a). This, however, does not contrast with the previous study with $Pr\gtrsim 1$ fluid, which suggests the wall mode drift frequency depends linearly on the reduced bifurcation parameter $\varepsilon = (Ra-Ra_w)/Ra_w$ \cite{ecke1992hopf,goldstein1993convection,zhang2020boundary,de2023robust}. In the limit of $Ek \rightarrow 0$, $\omega_{sw}/\Omega \propto Ek^2Pr^{-1}Ra = Ro_C^2$ \cite{de2023robust}, or by empirical relation $\omega_{sw}/\Omega \propto Ek^{5/3}Pr^{-4/3}Ra$ \cite{zhang2020boundary}. \Cref{fig:fwRoc} shows that with limited data, our wall mode frequency data also agree well with the linear scaling of the supercriticality, as observed in $Pr\gtrsim 1$ fluids. More comprehensive surveys are required for future studies to validate the applicability of linear $Ro_C$ scaling in liquid metal. 

\subsection{Comparison to large-scale vortices (LSV)}
The quadrupolar cyclone resembles many characteristics of Large-Scale Vortices (LSVs) observed in infinite plane \cite[e.g.,][]{stellmach2014approaching}. Owing to its quasi-2D nature, the LSV emerges from an inverse cascade of the kinetic energy to the largest available scale of the flow. Indeed, both quadrupolar cyclone and LSV are vertically coherent structures existing on a turbulent flow background. However, no laboratory evidence of LSV has been found experimentally. This prompts the question of what underlies the formation of the quadrupolar vortex observed here: is it solely driven by the development of the wall mode and its interaction with the bulk, or is the flow structure a result of energy transport to larger scales? Water experiments and simulations \cite{madonia2021velocimetry,de2023robust} point to wall mode as the source of this large-scale structure. Future study of the quadrupolar cyclone evolution with different aspect ratios is required to answer this question for the liquid metal RC systems, and perhaps can also help explain the difference in wall mode frequency and lateral-vertical velocity ratio between liquid metal and $Pr \gtrsim 1$ fluids.  

\section{Conclusions}
\label{conclusion}

We explored the dynamics of rotating Rayleigh-B\'enard convection (RC) in liquid gallium characterized by a low Prandtl number ($Pr = 0.027$) within a slender cylindrical container (aspect ratio $\Gamma = D/H = 1/2$). Employing a novel thermovelocimetric technique combining thermometry and Dopper velocimetry, we achieve simultaneous measurement of temperature and velocity fields, enabling a detailed examination of liquid metal convection patterns under rapid rotation. The study uncovers new insights into the complex interplay between geometry, rotational forces, and fluid properties in liquid metal RC inside a slender container. The results extend the understanding of wall-bulk interactions and their impact on flow structures and transport phenomena in incompressible low Prandtl number fluids, offering implications for theories and applications involving convective processes in liquid metals.

The formation of cyclonic structures under wall-bulk interaction is one of the key findings. Our thermovelocimetric approach highlights the emergence of a stable $m = 2$ cyclonic structure in the \textbf{RC60W} case, notable at low convective Rossby numbers. This structure, exhibiting a quadrupolar form, remains aligned with the cylinder's rotational axis and shows little variation along the cylinder's height. It appears that the slender geometry of the cylinder enhances wall modes which, in conjunction with the bulk flow, leads to the appearance of these large-scale vortices.

The dynamics of the interaction between wall and bulk flow seem to be significantly influenced by the convective Rossby number, $Ro_C$. At lower values, wall modes become more pronounced, driving the large-scale vortices through strong jet-like flows from the wall and influencing the azimuthal precession of the bulk flow. As the Rossby number increases, the quasi-2D flow structure will eventually be overwhelmed by the convective turbulence. 


Our findings not only align with previous studies conducted on fluids with $Pr \sim 1$ across different cylindrical aspect ratios but also demonstrate that in liquid metal, the interaction effects between the wall and the flow are markedly stronger. This suggests an enhanced role of the cylinder walls in heat and mass transport within low $Pr$ fluids. Future analysis of the fastest-growing bulk mode might be helpful in comparing it against the quadrupolar cyclone. Furthermore, the transition of quadrupolar cyclones into boundary zonal jets could be expected in higher $\Gamma$ liquid metal RC.




\section*{Acknowledgements}
We gratefully acknowledge the support of the NSF Geophysics Program (EAR awards \#1853196 and \#2143939) and the National Defense Science and Engineering Graduate Fellowship Program (J.A.A.). T.V. thanks Deutsche Forschungsgemeinschaft (DFG) under the grant VO 2331/3. Partial work conducted by Y.X. in this paper was supported by the U.S. Department of Energy under contract number DE-AC02-09CH11466, and under the Laboratory Directed Research and Development (LDRD) Program at Princeton Plasma Physics Laboratory.

\bibliographystyle{elsarticle-num} 
\bibliography{reference}






\clearpage
\appendix
\section{UDV profiles of RC12W and RC60W} 
\label{app1}

\begin{figure}[ht!]
	\centering 
	\includegraphics[width=\textwidth]{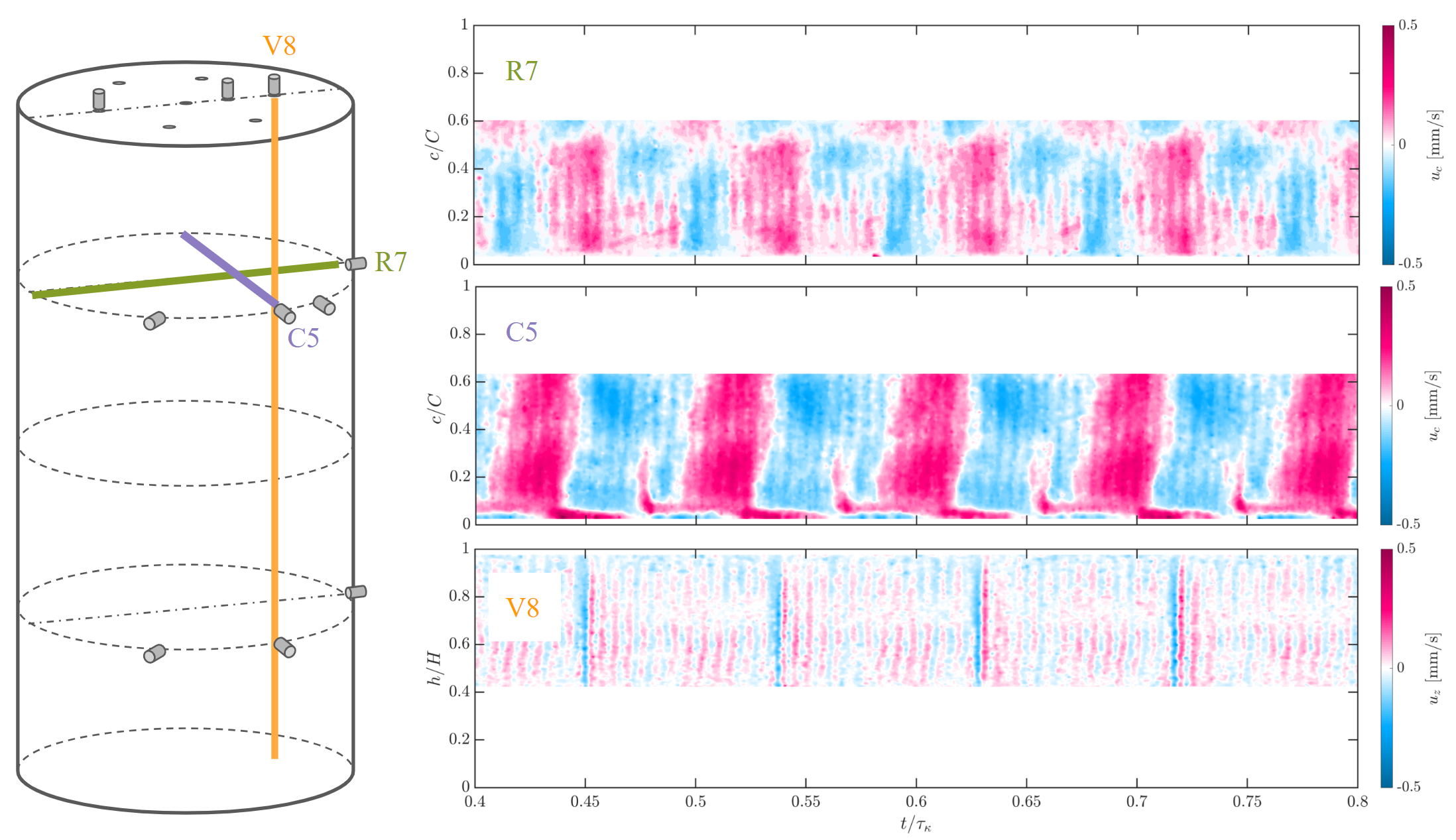}	
	\caption[]{UDV profiles of \textbf{RC12W} subject to a RH rotation. We obtained a partial velocity field along the beam due to the limitation of signal quality at a low-velocity regime (maximum velocity has an amplitude of less than $\approx 0.5 \, \mathrm{mm/s}$). Ripples (vertical stripes) of the velocity patterns at the smallest wavelength, although very weak, could be contributed by oscillatory mode at the onset. The small finger-shaped red structures between the red pattern in C5 indicate periodic inward jets coming from the sidewall. 
}\label{fig:12w}%
\end{figure}

\begin{figure}[ht!]
	\centering 
	\includegraphics[width=\textwidth]{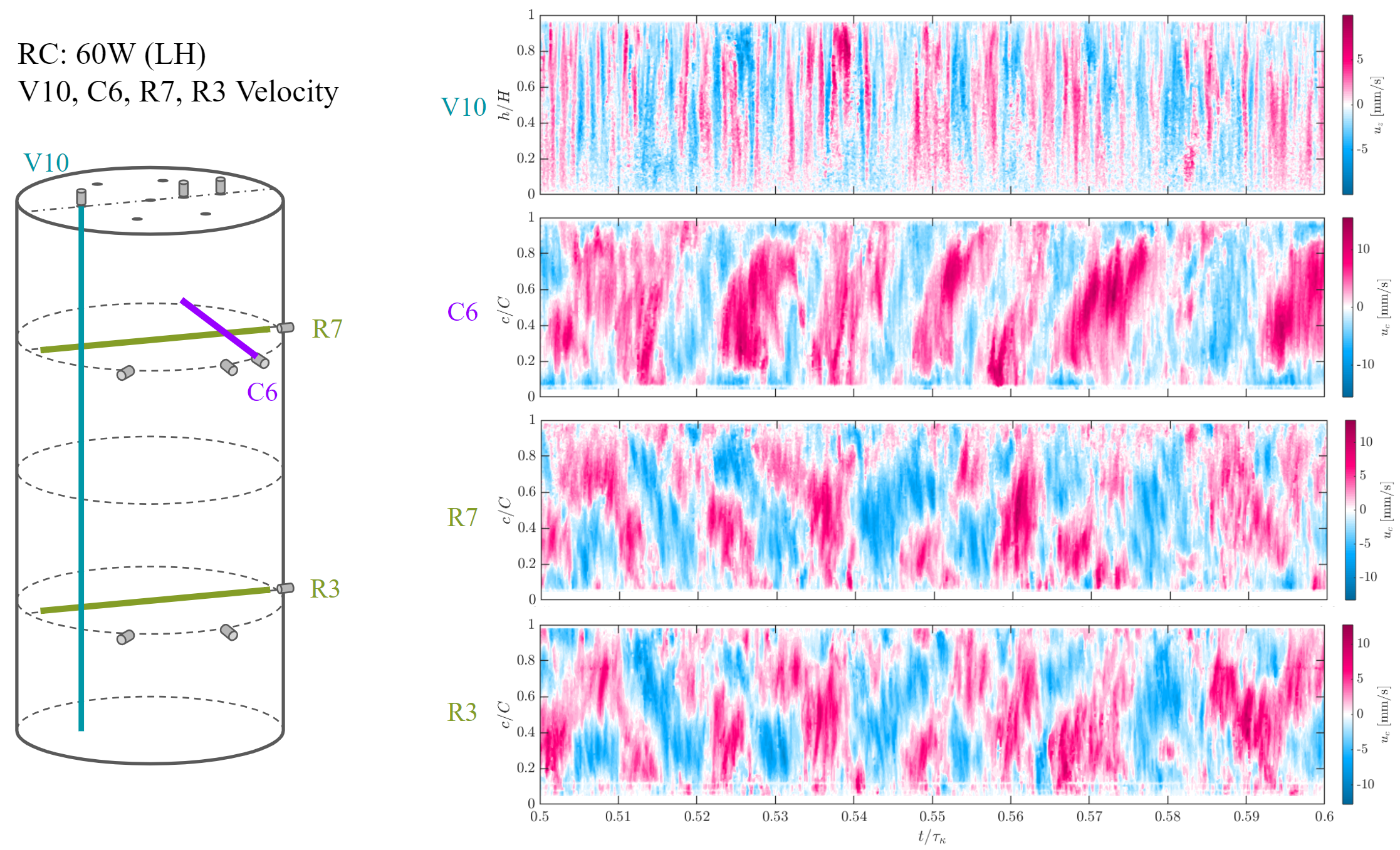}	
	\caption[]{Hovm\"oller diagrams of UDVs for the RC60W case. Four Doppler probes are set up as the schematics on the left. Probe R7 and R3 are the radial probes, V10 is the vertical probe, and C6 is the chord probe. The diagrams show the velocity in color along the beam direction. The vertical axes are the distance normalized by the path length. The horizontal axes are time normalized by thermal diffusion time $\tau_\kappa$. All the diagrams show $t= 0.1 \tau_\kappa$ intervals. For vertical probes, red represents positive velocity in the upward direction. For chord and radial probes, red represents velocity moving away. The maximum velocity for each probe's velocity profile is reported below: $u_{max} (V10) = 8.4\mathrm{mm/s}$, $u_{max} (C6) = 13.6\mathrm{mm/s}$, $u_{max} (R7) = 15.7\mathrm{mm/s}$, and $u_{max} (R3) = 14.9\mathrm{mm/s}$.
}\label{fig:dop60w}%
\end{figure}

\Cref{fig:12w} and \Cref{fig:dop60w} shows the setup and Hovm\"oller diagrams of UDV measurements. 
%
%

\clearpage

\section{Normal Modes}
\label{app2}
The incompressibility of the fluid guarantees that the velocity field is solenoidal and satisfies the continuity equation, $\boldsymbol{\nabla}\cdot \boldsymbol{u} = 0$. Alternative to the potential vortices model, we can assume the vortices are invariant in the vertical direction, we can define a velocity vector potential $\boldsymbol{\Psi}$ in $\hat z$, and has solutions as the Bessel function of the first kind in cylindrical coordinates,
\begin{equation}
    \boldsymbol{\Psi} = J_m (\alpha_{mn} r) \cos \left( m (\Phi - 2 \pi \tilde \omega t) \right),    
\end{equation}
where $\alpha_{mn}$ is the the $n$-th zero of the $m$-th Bessel function. Thus, the velocity field in a cylindrical coordinate has $\hat r$ and $\hat \phi$ components, 
\begin{equation}
\begin{split}
    \boldsymbol{u} &= \boldsymbol{\nabla} \times \boldsymbol{\Psi} \\
    &= J_m(\alpha_{mn} r) m \sin(m (\phi - 2 \pi \tilde \omega t)) \hat{r} \\
    &\quad +\left( J_m(\alpha_{mn} r) + r J'_m(\alpha_{mn} r) \alpha_{mn} \right) \cos(m (\phi - 2 \pi \tilde \omega t)) \hat{\phi}, 
\end{split}
\end{equation}
where $J'_m = \partial/\partial r J_m $. 

\begin{figure}[ht]
	\centering 
	\includegraphics[width=\textwidth]{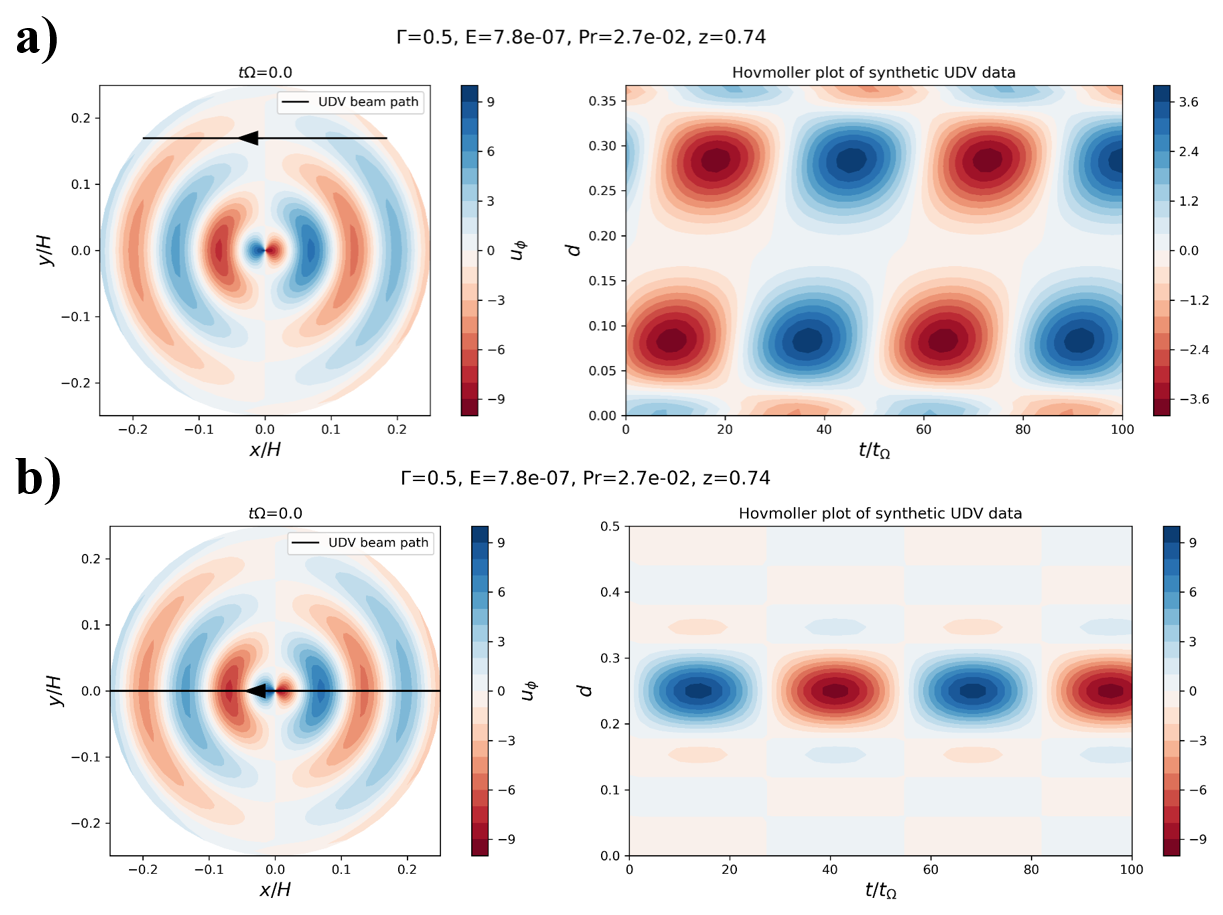}	
	\caption[]{Cross section of the velocity field and synthetic UDV data at a) C6 and b) R7 for oscillatory mode \cite{zhang_liao_2017} of \textbf{RC60W} case, at $Ek = 7.8\times10^{-7}$, $Pr = 2.7\times 10^{-3}$, $z = 0.74 H$.  
}\label{fig:osc}%
\end{figure}

\Cref{fig:osc} shows the onset prediction of the oscillatory modes and synthetic radial and chord Doppler data, using the solution by \cite{zhang_liao_2017}.

\clearpage
\section{Higher wavenumber models mismatch with observation} \label{app3}
\begin{figure}[ht]
	\centering 
	\includegraphics[width=0.8\textwidth]{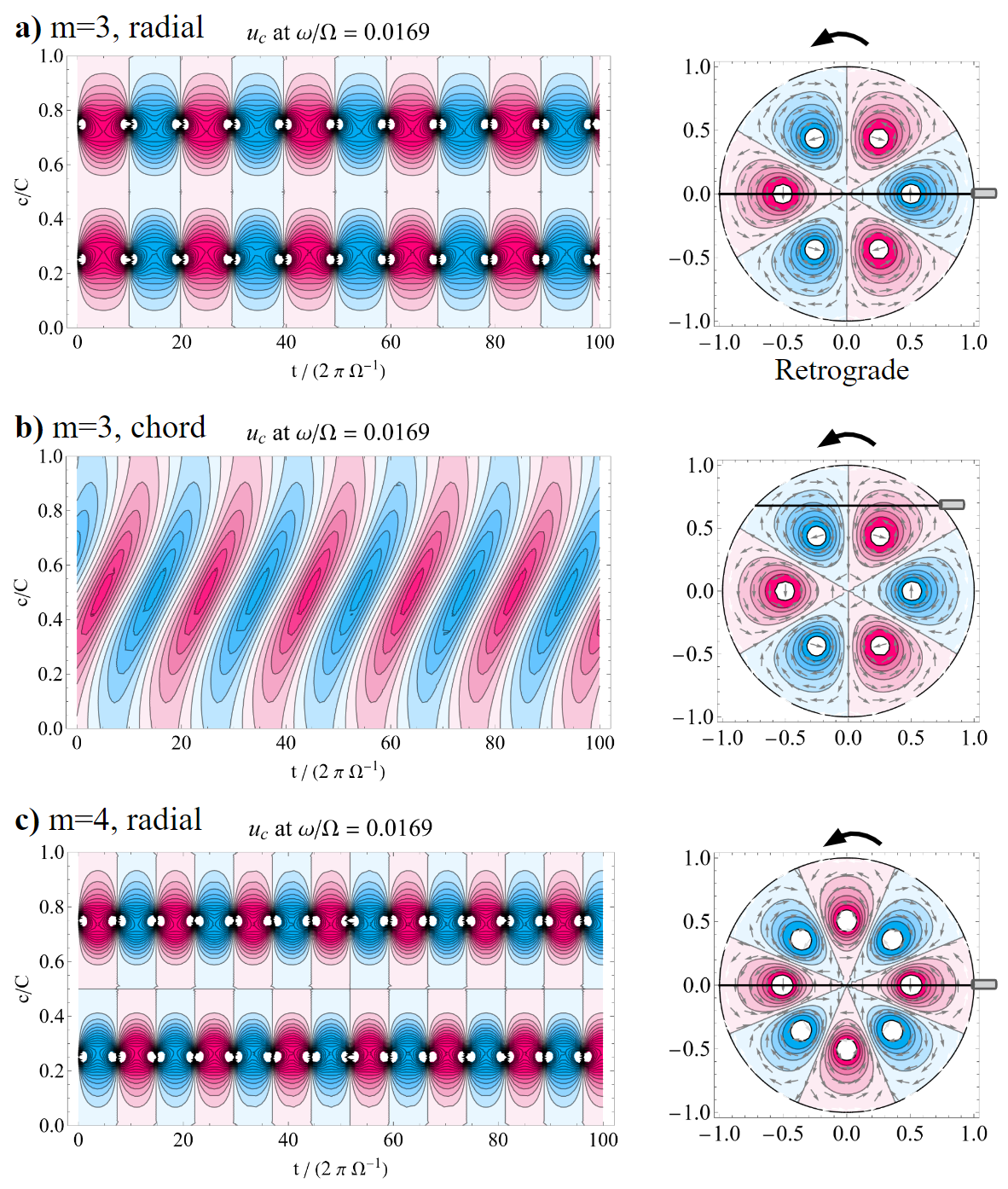}	
	\caption[]{ $m>2$ models of potential vortices for \textbf{RC60W} have higher frequencies than the experimental observation in \Cref{fig:synth6} b) and \Cref{fig:synth7} b) and different radial patterns if $m$ is odd.
}\label{fig:msynth}%
\end{figure}
\end{document}